\documentclass[a4paper,11pt]{article}
\usepackage{jheppub} 
\usepackage[T1]{fontenc} 
\usepackage{indentfirst}
\usepackage{subcaption}
\usepackage{mathtools}

\title{Gravitational Waves from Dark Yang-Mills Sectors}

\author[a,b]{James Halverson}
\author[c]{Cody Long}
\author[a]{Anindita Maiti}
\author[a,b]{Brent Nelson}
\author[a]{Gustavo Salinas}

\affiliation[a]{Department of Physics, Northeastern University \\ Boston, MA 02115, USA}
\affiliation[b]{The NSF AI Institute for Artificial Intelligence and Fundamental Interactions}
\affiliation[c]{Department of Physics and CMSA, Harvard University \\ Cambridge, MA 02138, USA}

\abstract{
    Dark Yang-Mills sectors, which are ubiquitous in the string landscape, may be reheated above their critical temperature and subsequently go through a confining first-order phase transition that produces stochastic gravitational waves in the early universe. Taking into account constraints from lattice and from Yang-Mills (center and Weyl) symmetries, we use a phenomenological model to construct an effective potential of the semi quark-gluon plasma phase, from which we compute the gravitational wave signal produced during confinement for numerous gauge groups. The signal is maximized when the dark sector dominates the energy density of the universe at the time of the phase transition. In that case, we find that it is within reach of the next-to-next generation of experiments (BBO, DECIGO) for a range of dark confinement scales near the weak scale.}

\begin{document} 
\maketitle
\flushbottom

\section{Introduction}
\label{sec:intro}

The first direct observations of gravitational waves (GWs) \cite{Abbott:2016blz, TheLIGOScientific:2017qsa, GBM:2017lvd} and the prospects for increasing experimental sensitivity in the next decades have put us on the precipice of a new era of multi-messenger astrophysics and cosmology. Gravitational wave experiments are poised not only to provide direct probes of energetic astrophysical phenomena, such as binary black hole mergers, but also provide a new window into the early universe via the measurement of a stochastic background of gravitational waves. Such a background may have a number of origins, including inflation, topological defects and cosmological first-order phase transitions (PTs) \cite{Allen:1996vm, Christensen:2018iqi}. In the visible sector, both electroweak (EW) gauge symmetry and the approximate chiral symmetry of QCD were spontaneously broken during phase transitions; these are, however, known not to be first-order\footnote{Some extensions of the SM change this, e.g., \cite{Iso:2017uuu, Weir_2018}.}.

Gravitational waves may also shed crucial light onto dark sectors. The existence of a dark matter component in our universe indicates that unknown particles might be hiding from observation. In optimistic scenarios, experiments may be sensitive to dark sectors that couple to the visible sector via portals that are not significantly suppressed. However, Nature may not be so forgiving: such portals may simply not exist, in which case the only interactions between the dark and visible sector are gravitational, a possibility that is unfortunately consistent with all current data. In this context, gravitational wave searches might be necessary to determine the properties of dark sectors.

In this paper we study stochastic gravitational waves produced during the confinement transition in pure Yang-Mills dark sectors. These are, of course, some of the simplest non-Abelian gauged dark sectors that might exist, but they are also well-motivated in string theory. For instance, dark gauge sectors naturally arise in the ten-dimensional $E_8\times E_8$ heterotic string itself \cite{PhysRevLett.54.502}, orbifold compactifications thereof \cite{Dixon:1985jw, Dixon:1986jc, IBANEZ198725, IBANEZ1988157, Lebedev_2007, Blaszczyk:2009in}, its free fermionic realizations \cite{ANTONIADIS198965, Faraggi:1997dc}, and it smooth Calabi-Yau compactifications \cite{Braun:2005ux, Bouchard:2005ag,Anderson:2012yf}; on $G_2$ compactifications of M-theory \cite{joyce1996, Acharya:1998pm, Halverson:2015vta}; and on seven-branes in F-theory. In fact, the latter provides the greatest evidence for dark gauge sectors: the three largest concrete F-theory ensembles \cite{Taylor:2015xtz, Halverson:2017ffz, Taylor:2017yqr}, which dwarf the rest of the currently known string landscape, all exhibits tens or hundreds of gauged dark sectors\footnote{In addition to motivating dark gauge sectors, these F-theory ensembles also motivate studies of axion-like particles, see e.g., \cite{Halverson:2019kna, Halverson:2019cmy}.}. In the simplest cases many of these factors are pure super-Yang-Mills sectors (from so-called non-Higgsable clusters) that flow to Yang-Mills sectors below the SUSY breaking scale; in the most common scenarios, the gauge groups are low rank $SU(N)$ groups, $G_2$, $F_4$, and $E_8$. Therefore in addition to being particularly simple extensions of the SM, dark Yang-Mills sectors are also well-motivated by ultraviolet considerations.

Pure Yang-Mills theories are expected to produce a first-order confining phase transition for almost all gauge groups of interest.\footnote{A notable exception being SU(2).} However, since there is not a first-principles description (other than lattice simulations) of the phase transition, many effective models have been considered in the literature. Among those are quasi-particle models \cite{PhysRevD.54.2399, PhysRevLett.94.172301, CASSING2007365, Castorina:2011ja}, approaches based on the functional renormalization group \cite{marhauser2008confinement, Braun:2010cy, Braun_2010, Herbst:2015ona}, Polyakov loop models \cite{Lo_2013, Hansen_2020}, as well as so-called matrix models \cite{Meisinger:2001cq,Dumitru:2010mj,Dumitru:2012fw, kondo2015confinementdeconfinement, Pisarski_2016, Nishimura_2018, Guo_2019, Korthals_Altes_2020, Hidaka:2020vna}. The latter are particularly interesting as they can be applied to any gauge group $\mathcal{G}$, the only input needed being the structure of the Lie algebra associated with $\mathcal{G}$.

The outline of this work is as follows. We first discuss, in Section \ref{sec:constraints}, the relevant symmetries for the construction of a matrix model of the confinement phase transition of pure Yang-Mills dark sectors along with results of lattice simulations that can be used to constrain it. We apply these considerations to construct the effective potential in concrete examples in Section \ref{sec:Vs}. The familiar case of SU($N$) is treated in detail and contrasted with the exceptional groups $G_2$ and $F_4$, also expected to confine in first-order phase transitions. Section \ref{sec:gws} then estimates the gravitational wave signal emitted during the confining transitions, accounting for theoretical uncertainties and determining their potential for detection in future experiments. We show that these transitions are not long-lasting, so that the GW signal emitted during the PT is suppressed, being only accessible to next-to-next generation searches. We end on Section \ref{sec:concl} with a summary and conclusions.

\section{Symmetry constraints and lattice}
\label{sec:constraints}

The confining phase transition in pure Yang-Mills theory can be described by an effective model based on the relevant order parameter, the Polyakov loop. In this section, we set the stage for the construction of a matrix model of (de)confinement in the absence of quarks, discussing the relevant symmetries as well as describing how lattice observables can be used to constrain the form of the effective potential.

\subsection{Symmetries of the effective potential}
\label{subsec:sym_V}

The order parameter for the confinement phase transition is the expectation value of the Polyakov loop $l$, the normalized trace of the thermal Wilson line $\mathbf{L}$, in the fundamental representation of the gauge group
\begin{equation}\label{eqn:polyloop}
\mathbf{L} = \mathcal{P} \exp \left( \mathrm{i} g \int_0^\beta A^a_0 (\vec x, \tau) T^a d\tau \right)~, ~~~ l = \frac{1}{d_f} {\rm Tr} \mathbf{L}~,
\end{equation}

\noindent with $g$ being the gauge coupling, $\beta$ the inverse of the temperature, $T^a$ the generators of $\mathfrak{g}$ in the fundamental and $d_f$ the dimension of the fundamental representation. Following the phenomenological approach of Refs. \cite{Meisinger:2001cq,Dumitru:2010mj,Dumitru:2012fw}, we consider an effective potential $V(\mathbf{L})$ for which the variables are the eigenvalues of the Wilson line $\mathbf{L}$, referred to as a \textit{matrix model} for confinement. This type of model can correctly describe the order of the phase transition for the gauge groups and, as we will see, allows for appropriate fits of thermodynamic observables studied on the lattice.

For simplicity, we take the time component of the vector potential to be constant\footnote{Note that this assumption implies $\langle L (\mathbf{A}_0) \rangle = L(\langle \mathbf{A}_0 \rangle)$, contradicting the non-saturation of the Jensen inequality $\langle L (\mathbf{A}_0) \rangle \leq L(\langle \mathbf{A}_0 \rangle)$ explicitly demonstrated in the functional renormalization approach (e.g., in \cite{marhauser2008confinement, Braun_2010, Braun:2010cy, Herbst:2015ona}). A less simplified model should be able to distinguish these two order parameters.} $\mathbf{A}_0 (\vec x, \tau) = A^a_0 (\vec x, \tau) T^a \equiv \mathbf{A}_0$. This component can always be diagonalized by a gauge transformation, so we take it to be an element of the Cartan subalgebra $\mathfrak{h}$ of the Lie algebra $\mathfrak{g}$ associated with the gauge group $\mathcal{G}$. The Cartan subalgebra is defined as the maximal subalgebra of mutually commuting generators. If $\{H_1, H_2, ..., H_r\}$ is a basis for $\mathfrak{h}$ (with $r$ being the rank of $\mathcal{G}$), a general element $H \in \mathfrak{h}$ can be written as $H = q_i H_i$, with $q_1, ... q_r$ being coordinates in the Cartan subalgebra.

Below the critical temperature\footnote{\label{foot}This temperature is of the same order as the confinement scale $\Lambda$ at which the running gauge coupling diverges. For example, in the case of SU($N$) one has $T_c \sim 1.5 \Lambda$ \cite{LUCINI2012279, Forestell:2016qhc}.} $T_{\rm c}$, the system is at a confined phase and the expectation value of the Polyakov loop vanishes identically\footnote{The fundamental Polyakov loop is related to the free energy $\mathcal{F}_{q \bar{q}/2}$ of a static quark-antiquark pair at infinite distance by $\langle l \rangle \sim \exp(-\beta \mathcal{F}_{q \bar{q}/2})$. In the confined state, $\mathcal{F}_{q \bar{q}/2} \rightarrow +\infty$, so $\langle l \rangle \rightarrow 0$.} $\langle l \rangle = 0$, while above $T_{\rm c}$ this order parameter becomes non-zero  $\langle l \rangle \neq 0$. Therefore, the effective potential has to be such that the (de)confinement phase transition is accompanied by spontaneous breakdown of center symmetry, so it should be invariant under center transformations. For SU($N$), center transformations are of the form
\begin{equation}
z_k = \exp (2 \pi \mathrm{i} k / N)
\end{equation}

\noindent at the Lie group level, with $k = 0, 1,..., N-1$. The thermal Wilson line in the fundamental transforms as $\mathbf{L} \rightarrow z_k \mathbf{L}$, so that such transformations act on the elements of the Cartan subalgebra as $H \rightarrow H + k ~ {\rm diag} (1,1,..., -(N-1) )/N$. Center symmetry is, however, absent for gauge groups with trivial centers, such as $G_2$, $F_4$ and $E_8$.

The roots $\alpha$ of the Lie algebra $\mathfrak{g}$ are linear functions on  the Cartan subalgebra $\mathfrak{h}$, defined by the commutation relations
\begin{equation}
[H, E_\alpha] = \alpha(H) E_\alpha \equiv \langle \alpha, H \rangle E_\alpha~,
\end{equation}

\noindent $E_\alpha$ being elements of $\mathfrak{g}$ ($E_\alpha \not\in \mathfrak{h}$ for non-zero roots) denoted root vectors. Note that in the first equality the roots are elements of the dual space $\mathfrak{h}^*$, but they can be mapped one-to-one into elements of $\mathfrak{h}$, as done in the second equality, if one takes $\alpha (\cdot) \equiv \langle \alpha, \cdot \rangle$, with  $\langle \cdot, \cdot \rangle$ denoting the Killing form. Any root can be written as a linear combination of the elements in a set $\Delta = \{\alpha_1,...,\alpha_r\}$ with integer coefficients that are all either non-negative or non-positive. The elements of $\Delta$ are called the positive simple roots of $\mathfrak{g}$.

With roots $\alpha$ as elements of the Cartan subalgebra, one can consider the group of reflections $w_{\alpha_i}$ about the hyperplanes perpendicular to each simple root $\alpha_i$, known as the Weyl group $W$. It can be shown that a Weyl transformation maps roots into roots, so that a particular choice of positive simple roots $\Delta$ can be mapped into any other choice $\Delta^\prime = w_\alpha \Delta$ by such reflections. Therefore, as all choices of $\Delta$ are equivalent, any function with domain in $\mathfrak{h}$, such as the effective potential we wish to construct, has to be invariant under the Weyl group.

Another important concept is that of the Weyl chamber, given by the set of $H \in \mathfrak{h}$ such that $\langle \alpha_i, H \rangle \geq 0$, for any positive simple root $\alpha_i$. This defines explicitly $\Delta$-dependent upper half planes in the Cartan subalgebra, so a Weyl transformation maps a Weyl chamber into another. It can be shown that the orbit under Weyl reflections of any point in the interior of a Weyl chamber ($\langle \alpha_i, H \rangle > 0$) has a number of elements equal to the order of the Weyl group. In other words, no Weyl transformation (other than the identity) maps a Weyl chamber to itself.

Writing\footnote{Here, we also denote by $H$ an element of the Cartan subalgebra in the fundamental representation, i.e., we make the identification $\Gamma(H) \equiv H$ with $\Gamma$ being the representation matrix in the fundamental. The only representation we use explicitly is the fundamental, so no ambiguities arise.} $\mathbf{A}_0 \equiv 2 \pi H/ \beta g$ in terms of $H \in \mathfrak{h}$, the thermal Wilson line becomes
\begin{equation}\label{eq:polyloop}
\mathbf{L} =\exp \left(2 \pi \mathrm{i} H \right)=\exp \left(2 \pi \mathrm{i} q_i H_i\right)~.
\end{equation}

\noindent We adopt a basis with elements $H_i$ such that their matrix exponential $\exp (2 \pi \mathrm{i} H_i)$ is either the identity or an element $z_k$ of the center for any $i = 1, ..., r$. With this choice, the effective potential becomes periodic in the $q$-coordinates with unit period and we identify $q_i \sim q_i + 1$ for each $i$, as elements connected by a center transformation should give the same value of the potential; its domain can then be restricted to the subset $0 \leq q_i < 1$.

Consider, as an illustration, the case of SU(3). The weights in the fundamental can be written as $(2 \alpha_1 + \alpha_2)/3$, $(- \alpha_1 + \alpha_2)/3$ and $-(\alpha_1 + 2\alpha_2)/3$ with $\alpha_1$ and $\alpha_2$ being the two positive simple roots. Then, a general element of the Cartan subalgebra is written, in the fundamental representation, as
\begin{equation}\label{eqn:su3_qs}
H = \frac{\alpha_1(H)}{3}{\rm diag} (2, -1, -1) +  \frac{\alpha_2(H)}{3}{\rm diag} (1, 1, -2)~.
\end{equation}

\noindent Taking $q_i = \alpha_i(H)$, $H_1 = {\rm diag} (2, -1, -1)/3$ and $H_2 = {\rm diag} (1, 1, -2)/3$ yields the desired form in Eq. \ref{eq:polyloop}. Note that this choice of $q$-coordinates implies that the interval $0 \leq q_i < 1$ is entirely contained within a single Weyl chamber, with its boundaries having at least one vanishing $q_i$. In the following, the choice of coordinates $q_i = \alpha_i(H)$ will be made for all gauge groups.

As we have argued above, the effective potential $V(q) \equiv V(q_1, q_2, ..., q_r)$ describing the confinement phase transition must be Weyl group-invariant. Weyl transformations generate permutations of \textit{all} roots, hence the potential has to be invariant under the corresponding permutations of its arguments. For example, any positive root of SU($N$) can be written as $\alpha = \sum_{i=m}^n \alpha_i$, with $1 \leq m \leq n \leq r$. As a consequence, the potential has to be invariant under permutations\footnote{To be clear, not all permutations of roots are generated by Weyl reflections, e.g., reflections do not change the angle between two roots. Thus, the symmetry imposed here is in fact larger than Weyl group invariance.} of the set $\{ \pm \sum_{i=m}^n q_i \}$, with $1 \leq m \leq n \leq r$.

\subsection{Thermodynamics of the gluon plasma}
\label{subsec:thermo}

Our goal is to construct an effective potential that describes the semi quark-gluon plasma (semi-QGP) in the absence of dynamical quarks, i.e., in pure Yang-Mills theories. The region of semi-QGP, which occurs in a range of temperatures from the critical temperature $T_{\rm c}$ to approximately $4 T_{\rm c}$, is characterized by a sharp increase of pressure starting from approximately zero in the confined phase (in units of the Stephan-Boltzmann limit, i.e., $p/p_{\rm SB} \approx 0$, with $p_{\rm SB}/T^4 \equiv d_A \pi^2/45$ and $d_A$ being the dimension of the adjoint representation) and asymptotically approaching the equation of state for an ideal gas at increasing temperatures.

We focus our attention on the region close to the critical temperature, looking for effective potentials that give the appropriate order for the phase transition and reproduce the behavior of thermodynamic quantities measured on the lattice. 


At high temperatures $T \gg T_{\rm c}$, the effective potential is given by the free energy of a gas of gluons in a constant background field $\mathbf{A_0}$ and can be found perturbatively as \cite{Dumitru:2012fw}
\begin{equation}\label{eqn:Vpt}
\frac{V_{\rm pt}}{T^4} = -\frac{p_{\rm SB}}{T^4} + \frac{2\pi^2}{3} \sum_\alpha B_4 (\langle \alpha, H \rangle)~,
\end{equation}

\noindent at one-loop order. The sum runs over all the roots $\alpha$ of $\mathfrak{g}$ and the function $B_4(x)$ is a shifted Bernoulli polynomial
\begin{equation}\label{eqn:B4}
B_4(x) = - \frac{3}{\pi^4}\left(\sum_{n=1}^\infty \frac{1}{n^4}\mathrm{e}^{2\pi \mathrm{i} n x} -\sum_{n=1}^\infty \frac{1}{n^4}\right)  = (x-\lfloor x \rfloor)^2 [1- (x-\lfloor x \rfloor)]^2~,
\end{equation}

\noindent with $\lfloor \cdot \rfloor$ being the floor function. We take the effective potential in the semi-QGP region to be the sum of the perturbative contribution $V_{\rm pt}$ in Eq. (\ref{eqn:Vpt}) and a nonperturbative contribution $V_{\rm npt}$, that respects the symmetries discussed in Section \ref{subsec:sym_V}. Note also that on the interval $0 < x < 1$ the function $B_4$ is polynomial; it is, however, not analytic at the origin (its third derivative involves the divergent sum $\sum_{n=1}^\infty  n^{-1}$) nor at any integer value of $x$. Therefore, the perturbative part of the effective potential is polynomial in the interior of a Weyl chamber\footnote{Note that we also need the restriction $\langle \alpha, H \rangle <1$ for any root $\alpha$. Henceforth, we use the term Weyl chamber to describe the region defined by the set of inequalities $0 \leq \langle \alpha, H \rangle  \leq 1$, with $\alpha$ being any positive root.}, but has singular behavior on the hyperplanes perpendicular to the roots, i.e., at the boundaries of the Weyl chambers. To avoid introducing additional singularities, we assume that the nonperturbative part of the potential is also polynomial in the interior of the Weyl chambers. As we will see, this assumption leads to Bernoulli polynomials of all (even) orders as building blocks of $V_{\rm npt}$. In particular, the shifted Bernoulli polynomial of degree two, given by $B_2(x) = (x-\lfloor x \rfloor)[1-(x-\lfloor x \rfloor)] = [B_4(x)]^{1/2} $ will be used extensively.

\subsection{Lattice observables}
\label{subsec:lattice}

The nature of the confinement phase transition, either continuous or not, can be determined on the lattice from the behavior of the order parameter $\langle l \rangle$ at the transition temperature. We are interested in first-order PTs, as these can potentially produce significant stochastic gravitational wave signals \cite{Schwaller:2015tja}. This type of transition involves a discontinuous change in the Polyakov loop at $T = T_{\rm c}$. Lattice simulations of pure SU($N$) gluodynamics have determined that the confinement phase transition is indeed first-order for $N \geq 3$ colors\footnote{These calculations were performed only up to $N = 8$. They, however, show that the first-order transition gets stronger with increasing $N$; one then expects that the transition continues to be of first order for arbitrary values of $N \geq 3$.} \cite{Lucini:2002ku, Lucini:2003zr, Lucini_2004, Lucini_2005}. Similar behavior was also found for gauge groups of the Sp($N$) type \cite{Holland:2003kg}. In addition, the phase transition for the exceptional group $G_2$ was shown to be discontinuous (see, e.g., \cite{Pepe:2005sz, Pepe:2006er, Bruno:2014rxa}), with $\langle l \rangle \approx 0$ below $T_{\rm c}$, even in the absence of center symmetry.

For gauge groups with a large number of gluons, there is a large mismatch between the number of degrees of freedom above (gluons) and below (color singlet glueballs) the critical temperature, as the latter is essentially independent of the dimension of the group. As such, one can expect, as conjectured in \cite{Pepe:2004rc}, the confinement phase transition to also be of first order in the case of larger gauge groups, such as $F_4$ and $E_8$, not yet studied on the lattice.

\begin{figure}
\makebox[1\textwidth][c]{
\begin{subfigure}{0.6\textwidth}
    \centering\includegraphics[scale=0.65]{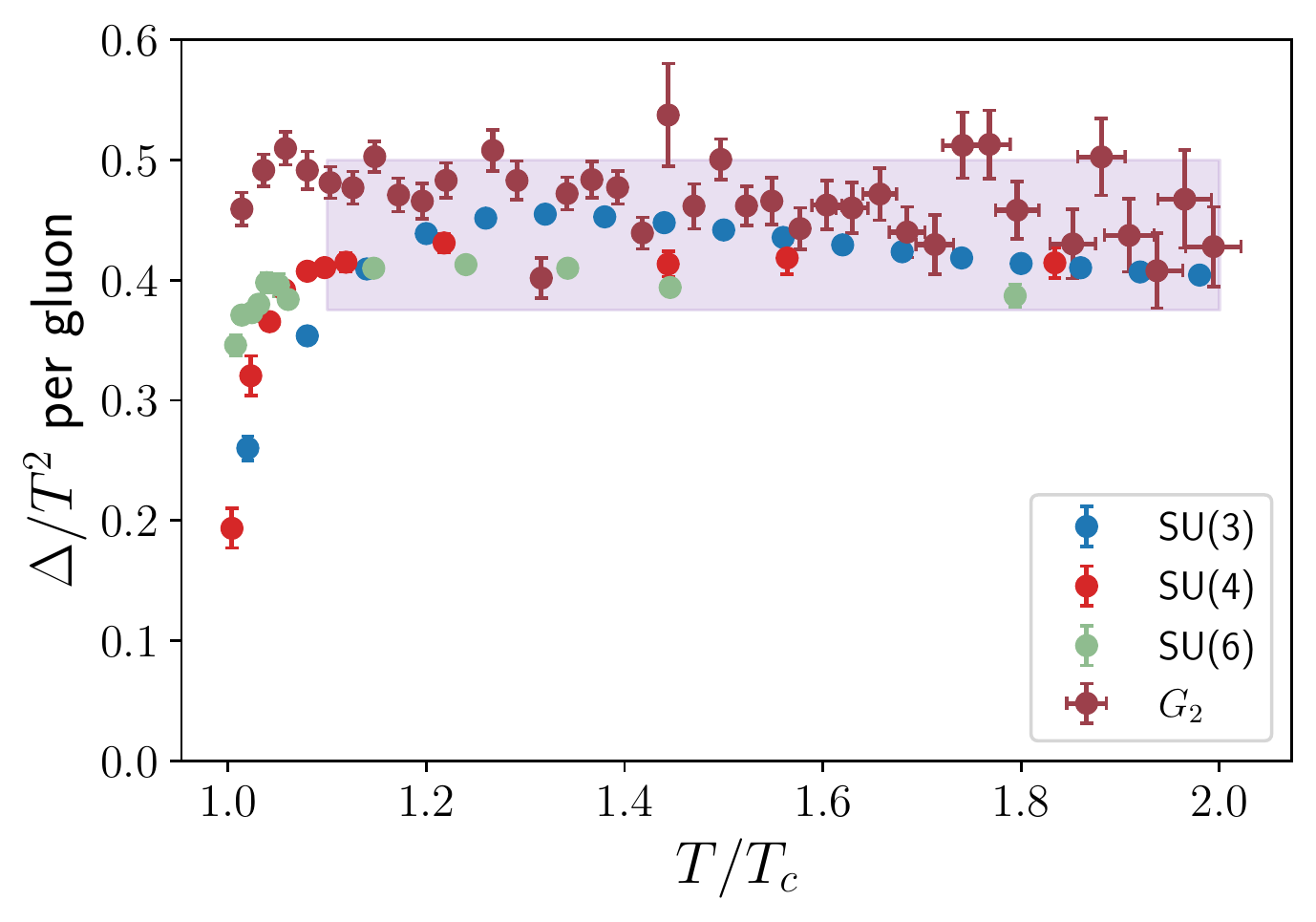}
    \caption[a]{} \label{fig:deltas}
\end{subfigure}
\begin{subfigure}{0.6\textwidth}
    \centering\includegraphics[scale=0.65]{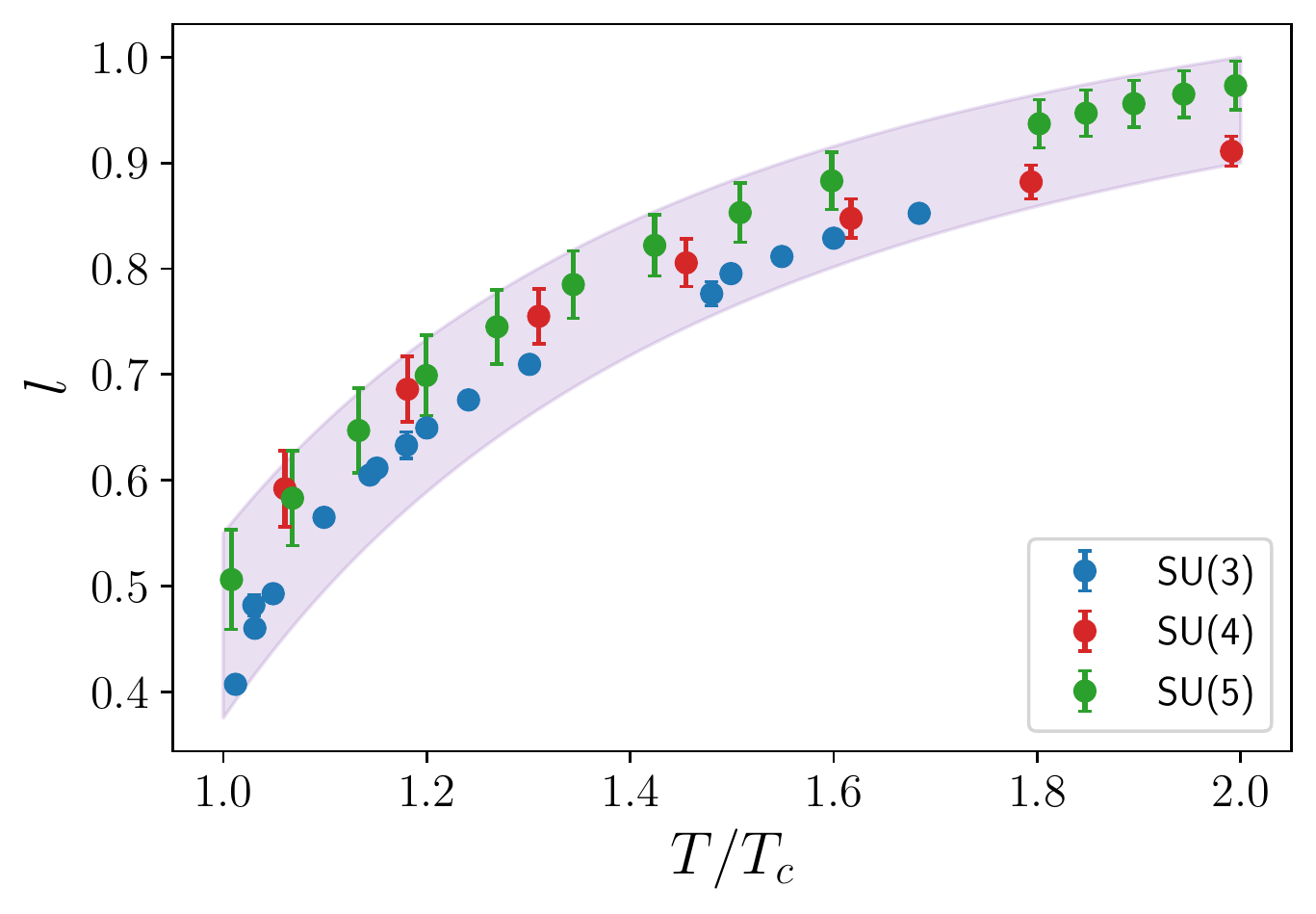}
    \caption{} \label{fig:ls}
\end{subfigure}
}
\caption{The behavior of thermodynamic quantities on the lattice; the data for the interaction measure (i) is taken from \cite{Caselle:2018kap} for SU(3), from \cite{Datta:2010sq, Panero:2009tv} for SU(4) and SU(6) (the figure only shows data points from \cite{Datta:2010sq} for clarity) and for the exceptional group $G_2$ adapted from \cite{Bruno:2014rxa}. The data for the renormalized Polyakov loop (ii) of SU($N$) is reproduced from Ref. \cite{Gupta:2007ax} for $N=3$, and \cite{Mykkanen:2012ri} for $N=4$ and $N=5$.}
\end{figure}

Lattice simulations also seem to indicate the temperature dependence of the nonperturbative part of the effective potential. This can most clearly be seen in the behavior of the interaction measure $\Delta$, defined as
\begin{equation}\label{eqn:trace_anom}
\frac{\Delta}{T^2} \equiv \frac{e - 3 p}{T^2} =  - T^3 \frac{\partial}{\partial T} \frac{V(q_{\rm min}(T);T)}{T^4}~,
\end{equation}

\noindent where $e(T)$ is the energy density and $p(T) = - V(q_{\rm min}(T);T)$ the pressure of the gas of gluons with $q_{\rm min}(T)$ denoting the coordinates of the global minimum of the potential at a temperature $T$. Above the phase transition, in the interval $1.1 T_{\rm c} \lesssim T \lesssim 4 T_{\rm c}$, the interaction measure is observed to be directly proportional to $T^2$ for all the gauge groups studied on the lattice\footnote{The case of SU($2$), which displays a second-order phase transition, appears to slightly deviate from this behavior \cite{Giudice:2017dor}.}, as shown in Fig. \ref{fig:deltas} for the groups SU($N$) with $N = 3,4,6$ and $G_2$. In this interval, the interaction measure divided by the square of the temperature is approximately constant with its value per gluon being approximately the same for each group: all data points fall (within error) in the range $0.38 \lesssim \Delta/d_A T^2 \lesssim 0.5$. In the following, we refer to this region as the \textit{expected region}, shaded in purple. This behavior indicates that the dominant contribution to $V_{\rm npt}$ should be, according to Eq. (\ref{eqn:trace_anom}), proportional to $T^2$, at least for temperatures right above the PT. Following Ref. \citep{Dumitru:2012fw}, we also allow for a temperature independent constant. In addition, we continue the expansion in even power of the temperature and include a term proportional to $T^{-2}$ with coefficient independent of the coordinates $q$. As we will see, this extra term allows for a better fit of our model to lattice thermodynamics results.

For SU($N$) groups, measurements of the renormalized Polyakov loop are available from the lattice in the cases $N=3$ \cite{Gupta:2007ax}, $N=4$ and $N=5$ \cite{Mykkanen:2012ri}. These are shown in Fig. \ref{fig:ls}. The data points show a similar trend for the different number of colors shown. Thus, we again select an expected region for the value of the Polyakov loop, shown in purple, and make the assumption that the renormalized Polyakov loop approximately falls within this region for an arbitrary number of colors, as well as for other gauge groups. It is fitted by $l(q_{\rm min})$ in our model. In addition, the latent heat for the SU($N$) transition was determined in Ref. \cite{Datta:2010sq} and can be used to further constrain our effective model for SU($N$), being given by the discontinuity in the interaction measure across the transition $\delta e = \Delta(T_{\rm c}^+) - \Delta(T_{\rm c}^-)$. For other gauge groups, we do not impose a value for the latent heat as a constraint, as these are not yet available from lattice studies.

The approximately universal behavior of the interaction measure and of the renormalized Polyakov loop described in this section will be used as a guide for our effective description of the gluon plasma close to the critical temperature. In the next section, we combine the symmetry considerations of Section \ref{subsec:sym_V} with these lattice results to construct an effective model for the confinement phase transition. We focus on the interval $T_{\rm c} \lesssim T \lesssim 2 T_{\rm c}$, since, as explained in the following, this allows for our simplified model to adequately fit the necessary observables.

\section{The effective potential}
\label{sec:Vs}

Given the symmetry and lattice constraints introduced in Section \ref{sec:constraints}, in this Section we construct the effective potentials describing the semi quark-gluon plasma phase that characterizes Yang-Mills theories just above $T_c$. Specifically, we impose center and Weyl group invariance as well as the expectations for thermodynamic quantities such as the interaction measure and for the Polyakov loop inspired by the apparent universality of lattice results discussed above. In Section \ref{subsec:suN}, we treat the more familiar case of SU($N$) and contrast it with the exceptional cases of $G_2$ and $F_4$ in Section \ref{sub:g2f4}.

\subsection{SU($N$)}
\label{subsec:suN}

We start by generalizing the choice in Eq. (\ref{eqn:su3_qs}) of coordinates  in the Cartan subalgebra of SU(3) to an arbitrary number of colors $N \geq 3$. The positive simple roots of the Lie algebra can be written as $\alpha_i = \mu_i - \mu_{i+1}$, with $\mu_i$ ($i = 1, ..., N-1$) being the weights in the fundamental representation and $\mu_N \equiv -(\mu_1 + \mu_2 + ... + \mu_{N-1})$. Inverting these relations, we obtain
\begin{equation}\label{eqn:suN_mus}
\left(\begin{array}{c}
\mu_1 \\
\mu_2 \\
\mu_3 \\
\vdots \\
\mu_{N-1}
\end{array}\right) = \frac{1}{N} \left(\begin{array}{cccccc}
N-1 & N-2 & N-3 & & 2 & 1 \\
-1 & N-2 & N-3 & & 2 & 1 \\ 
-1 & -2 & N-3 & & 2 & 1 \\ 
-1 & -2 & -3 & & 2 & 1 \\ 
& & & \ddots & & \\
-1 & -2 & -3 & & -(N-2) & 1 \\ 
\end{array}\right) \left(\begin{array}{c}
\alpha_1 \\
\alpha_2 \\
\alpha_3 \\
\vdots \\
\alpha_{N-1}
\end{array}\right)
\end{equation}

\noindent along with $\mu_N = (-\alpha_1 - 2\alpha_2 - ... - (N-1) \alpha_{N-1} )/N$. Thus, a general element $H \in \mathfrak{h}$ can be written as
\begin{equation}
H = \frac{\alpha_1(H)}{N} {\rm diag} (N-1, -1, ..., -1) + \frac{\alpha_2(H)}{N} {\rm diag} (N-2, N-2, ..., -2) + ... \equiv q_i H_i~, 
\end{equation}

\noindent with the coordinates again chosen as $q_i \equiv \alpha_i(H)$.

In terms of these coordinates, the perturbative part of the effective potential from Eq. (\ref{eqn:Vpt}) can be written as
\begin{equation}\label{eqn:suN_Vpt}
\frac{V_{\rm pt}}{T^4} = -\frac{(N^2-1) \pi^2}{45} + \frac{2\pi^2}{3} \sum_{q_\alpha} B_4 (q_\alpha)~,
\end{equation}

\noindent with $q_\alpha$ belonging to the set $\{ \pm \sum_{i=m}^n q_i \}$, with $1 \leq m \leq n \leq N-1$. It is explicitly invariant under the Weyl group, as the summatory runs over all roots of $\mathfrak{g}$. We can also check invariance under center symmetry. It was shown in Section \ref{subsec:sym_V} that a center transformation $z_k$ acts on the weights of the fundamental as $\mu_i \rightarrow \mu_i + k/N$ for $i = 1, ..., N-1$ and $\mu_N \rightarrow \mu_N - k (N-1)/N$. One can check, using Eq. (\ref{eqn:suN_mus}), that this transformation shifts the roots $\alpha_i$ by an integer. The function $B_4 (x)$ in Eq. (\ref{eqn:B4}) has unit period and the change in its argument under a center transformation is an integer for each element in the sum, so $V_{\rm pt}(q)$ is indeed invariant under center symmetry. This agrees with the fact that the adjoint representation has zero $N$-ality, so that terms constructed from the adjoint Polyakov loop (involving all the roots) should be left invariant by center transformations.

We now consider the nonperturbative contribution to the potential. Combining the symmetries from Section \ref{subsec:sym_V} with the lattice results discussed in Section \ref{subsec:lattice}, we assume that it is a Weyl group and center-invariant almost-everywhere polynomial function, with a dominant component proportional to $T^2$. By almost-everywhere polynomial we mean a function that, like the Bernoulli polynomial in (\ref{eqn:B4}), is polynomial except at the boundaries of Weyl chambers. For simplicity, we assume a polynomial of degree four, this being the lowest degree necessary to describe a first-order phase transition as a thermal transition from a metastable vacuum to the true vacuum of the theory, separated by a barrier of finite height\footnote{This requires the implicit assumption that, at temperatures close to the phase transition, each quantum state is dominated by a single value of $\mathbf{A_0}$. This assumption is often incorrect; for example, it cannot explain the small value of the adjoint SU($N$) Polyakov loop in the confined phase, seen on the lattice \cite{Gupta:2007ax}. However, the thermodynamic
behavior of the confinement phase transition can still be modeled properly. As explained in \cite{Dumitru:2012fw}, such a treatment can be seen as a lowest order approximation of a type of large-$N$ expansion.}. To account for Weyl symmetry, we consider terms of the form
\begin{equation}\label{eqn:general_terms}
\sum_{\alpha \in W \cdot \tilde{\alpha}} P_1[ \alpha(H) ]~, \sum_{\substack{\alpha, \alpha^\prime \in W \cdot \tilde{\alpha}\\ \alpha \neq \alpha^\prime}} P_2[ \alpha(H), \alpha^\prime(H) ]~, ~~...
\end{equation}

\noindent where $W \cdot \tilde{\alpha}$ denotes the orbit of a root $\tilde{\alpha}$ under Weyl transformations and $P_1, P_2, ...$ are polynomials in the interior of a Weyl chamber of degree (less or equal to) four . For $\mathfrak{su}$($N$), this orbit is the set of all roots. For other algebras, roots might have different lengths and, as we will see, one has to include terms summing over distinct orbits.

The periodicity of the coordinates in the Cartan subalgebra constrains the form of the polynomials $P$. Consider, for example, a term containing one of the coordinates, $q_i$. If one performs a Weyl reflection $w_{\alpha_i}$ that takes $q_i$ into $-q_i$, followed by the transformation $q_i \rightarrow q_i + 1$, the potential should be left invariant. Note that the resulting transformation, $q_i \rightarrow 1 - q_i$, keeps the coordinate within its restricted domain $0 \leq q_i < 1$. Thus, the polynomials $P$ should have the property $P(1-q) = P(q)$. Bernoulli polynomials of degree $n$ obey $B_n(1-q) = (-1)^n B_n(q)$, so the ones with even degree form the appropriate basis for our construction. Therefore, the most general terms of the form (\ref{eqn:general_terms}) obeying these symmetries are
\begin{equation}\label{eqn:VsuN_terms}
V_1 = \frac{1}{2}\sum_\alpha B_2(\alpha) ~, ~ V_2 = \frac{1}{8} \sum_{\alpha \neq \alpha^\prime} B_2(\alpha) B_2(\alpha^\prime) ~, ~ V_3 = \frac{1}{2} \sum_{\alpha} B_4(\alpha) ~,
\end{equation}

\noindent where the multiplying factors are chosen for convenience and the sums run over all roots of $\mathfrak{su}$($N$). For clarity, we write these terms explicitly in the case of SU(3),
\begin{align}
\begin{split}
V^{\rm SU(3)}_1 &= q_1 (1-q_1) + q_2 (1-q_2) + q_3(1-q_3) \\
V^{\rm SU(3)}_2 &= q_1 (1-q_1) q_2 (1-q_2) + q_1 (1-q_1) q_3 (1-q_3) + q_2 (1-q_2) q_3 (1-q_3) \\
V^{\rm SU(3)}_3 &= q_1^2 (1-q_1)^2 + q_2^2 (1-q_2)^2 + q_3^2(1-q_3)^2~,
\end{split}
\end{align}

\noindent with $q_3 \equiv q_1 + q_2$. Note that, since $V_1$ is of degree two, a term proportional to $V_1^2$ is also allowed by the symmetries. Such a term is, however, a linear combination of $V_2$ and $V_3$ above. The nonperturbative part of the effective potential is then taken to be of the form
\begin{equation}\label{eqn:suN_Vnpt}
V_{\rm npt}(q) = T_{\rm c}^2 T^2 \left(c_0 + \sum_{i = 1}^3 c_i V_i (q) \right) + d_1 T_{\rm c}^4 + d_2 \frac{T_{\rm c}^6}{T^2}~,
\end{equation}

\noindent where $c_i, d_j$ are coefficients still to be set. As explained in Section \ref{subsec:lattice}, the temperature dependence of $V_{\rm npt}(q)$ right above the critical temperature is mainly given by a component proportional to $T^2$ and we write it as a linear combination of terms in Eq. (\ref{eqn:VsuN_terms}). As these are the only $q$-dependent terms, it encodes the dynamics of the phase transition. As adopted in Ref. \cite{Dumitru:2012fw}, we allow for a temperature-independent constant\footnote{Note, however, that the authors of that work do not allow for a $V_2$ term.} $d_1 T_{\rm c}^4$, where the factor $T_{\rm c}^4$ makes the coefficient $d_1$ dimensionless. 

Close to the critical temperature, it is reasonable to allow for some physics to give increasing contributions to the effective potential for decreasing $T$, corresponding to the appearance of terms proportional to negative powers of $T$; if such physics does not exist, lattice will fit the coefficients to zero. Thus, we continue the expansion in even powers of the temperature and add a term proportional to $T^{-2}$. As we will see, the inclusion of such a term allows for a correct description of the evolution of the Polyakov loop as a function of temperature, while simultaneously fitting other thermodynamic quantities. The model, however, fails to do so if one sets $d_2 = 0$.

The confined state, the center-symmetric state with vanishing Polyakov loop, has coordinates $(q_c)_i = 1/N$ for all $i = 1, ..., N-1$, which amounts to having all the eigenvalues of the thermal Wilson line $\mathbf{L}$ equally spaced along the unit circle. As observed on the lattice, the confinement transition at $T_{\rm c}$ does not take the system directly to the perturbative vacuum $(q_d)_i = 0$ (i.e., $\langle l(T_{\rm c}^+) \rangle \neq 1$). Thus, the discontinuous transition happens between a metastable state at $q_c$ and another state $q_t \neq q_d$ inside the Weyl chamber. The coefficients in Eq. (\ref{eqn:suN_Vnpt}) are not all independent. First, it is necessary to impose that the phase transition happens at $T = T_{\rm c}$. In addition, we assume that the pressure of the glueball gas in the confined state vanishes\footnote{It is certainly true that the pressure in the confined state is always much smaller than at much higher temperatures, but a nonvanishing value can be measured on the lattice (see e.g, \cite{Borsanyi:2012ve}). Its value, however, is small enough that this assumption should not change our results appreciably.}, $V(q_c; T_{\rm c}) = 0$.


Before explicitly imposing the constraints discussed above, let us comment on a simplifying assumption, termed the \textit{uniform eigenvalue ansatz}, i.e., the assumption that the eigenvalues for the thermal Wilson line $\mathbf{L}$ on the minimum of the effective potential are equally displaced along a section of the unit circle, for all temperatures. This amounts to taking $q_i (T) = (1 - r(T))/N$ for any $i$, with $r(T \leq T_{\rm c}) = 0$ at the confined state and $r(T \rightarrow \infty) = 1$ at the perturbative vacuum. This ansatz reduces the problem to a one-dimensional thermal transition between two vacua\footnote{This ansatz is, however, only \textit{approximately} realized in the exact solution.} and it will be used for large numbers of colors, $N \geq 8$, allowing for an estimation of the thermal transition rate in those cases. Under this assumption, the terms defined in (\ref{eqn:VsuN_terms}) become
\begin{align}
\begin{split}\label{eqn:VsuN_terms_r}
V_1 &= \frac{N^2-1}{12} (1-r^2) \\
V_3 &= \frac{N^2-1}{60} \left[ 1 + \frac{1}{N^2} - r^2 \left(1 + \frac{6}{N^2}\right) - 2 r^3 \left(1 - \frac{4}{N^2}\right) + 2 r^4 \left(1 - \frac{3}{2 N^2}\right)\right]\\
V_2 &= \frac{1}{2}(V_1^2 - V_3)~.
\end{split}
\end{align}

\noindent Note that the uniform eigenvalue ansatz amounts to having the global minimum of the potential always located at the line that is equidistant from the faces at the boundary of the Weyl chamber. In the next section, we use this fact to generalize the uniform eigenvalue ansatz.

As explained in Section \ref{subsec:lattice}, Polyakov loop data constrains the form of the effective potential. In particular, we chose the state $q_t$ above the transition in such a way that it matches the value for $\langle l \rangle$ observed on the lattice (e.g., $l(T_{\rm c}^+) \approx 0.4$ for SU(3) and $l(T_{\rm c}^+) \approx 0.5$ for SU(4)). For larger numbers of colors, we fit the Polyakov loop in our model to match the mid-sectional curve on the expected region shown in Fig. \ref{fig:ls}.

Finally, we impose that the latent heat of the transition agrees with the values found in Ref. \cite{Datta:2010sq},
\begin{equation}\label{eqn:latheat}
\Delta(T_{\rm c}^+) - \Delta(T_{\rm c}^-) \approx \Delta(T_{\rm c}^+) = (N^2-1)T_{\rm c}^4 \left(0.388-\frac{1.61}{N^2} \right)~,
\end{equation}

\noindent where we made the approximation $\Delta(T_{\rm c}^-) \equiv e(T_{\rm c}^-) - 3p(T_{\rm c}^-) \approx 0$, as both the energy density $e$ and the pressure $p$ are negligible in the confined phase. When applied to the nonperturbative potential in Eq. (\ref{eqn:suN_Vnpt}), these constraints, along with the ones discussed previously, reduce the number of independent coefficients from six to two. The remaining coefficients are then found by fitting to lattice data, the results for $N = 3,4,6$ being shown in Fig. \ref{fig:fitsuN}. The numerical values for all coefficients in Eq. (\ref{eqn:suN_Vnpt}) are shown in the Appendix. It is clear that the model can quantitatively describe lattice thermodynamics in the interval of interest, from $T_{\rm c}$ up to approximately $2 T_{\rm c}$. For temperatures not in this range, our model gives wildly incorrect or even unphysical results (e.g., negative pressure). This is, of course, a result of trying to describe a strongly interacting system with a potential that can be nicely written down as a sum of a non-perturbative and a one-loop order perturbative term. Thus, we content ourselves with the less ambitious goal of trying to model the region close to the PT.

\begin{figure}
\makebox[1\textwidth][c]{
\begin{subfigure}{0.6\textwidth}
    \centering\includegraphics[scale=0.65]{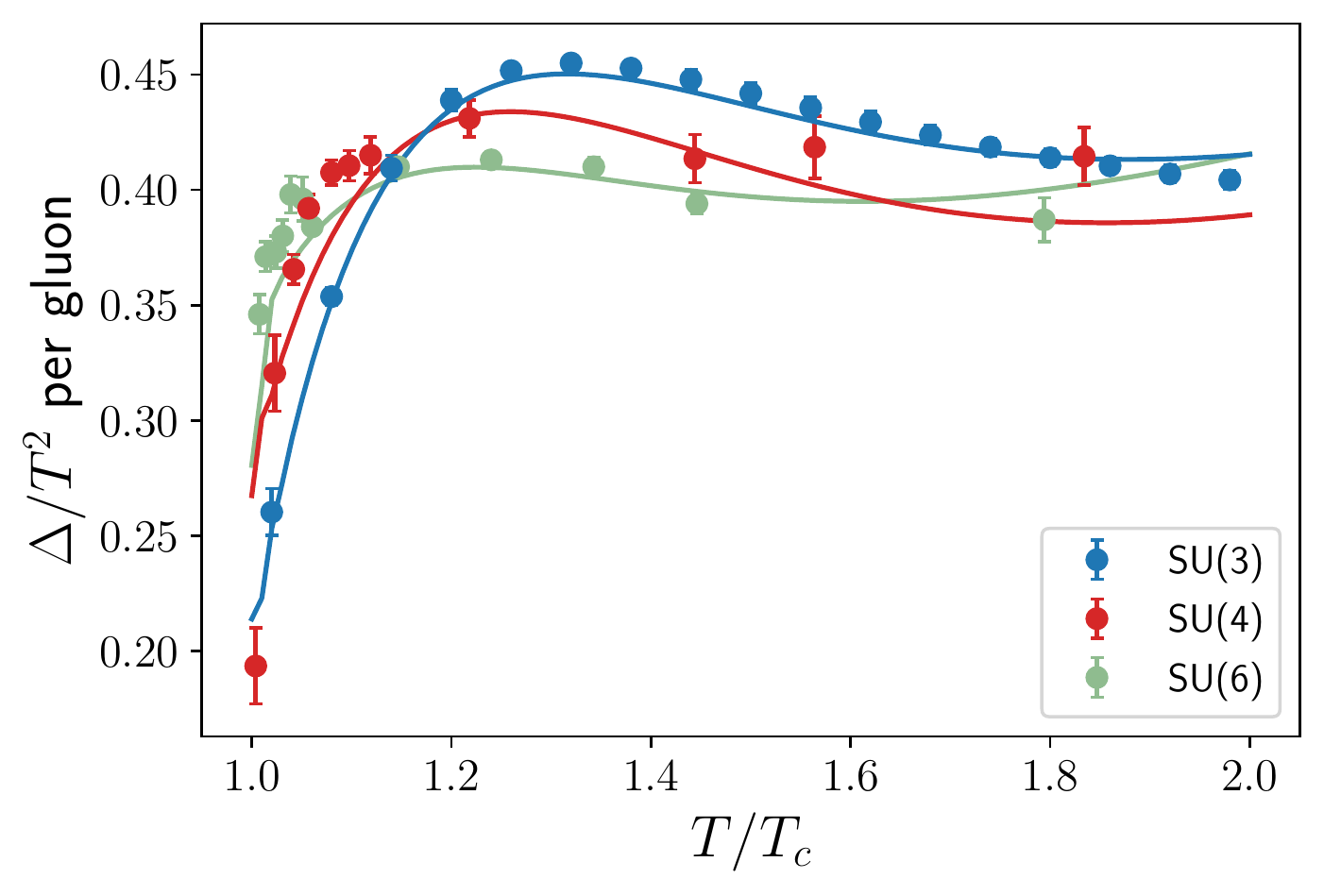}
    \caption{} \label{fig:fitsuN2}
\end{subfigure}
\begin{subfigure}{0.6\textwidth}
    \centering\includegraphics[scale=0.65]{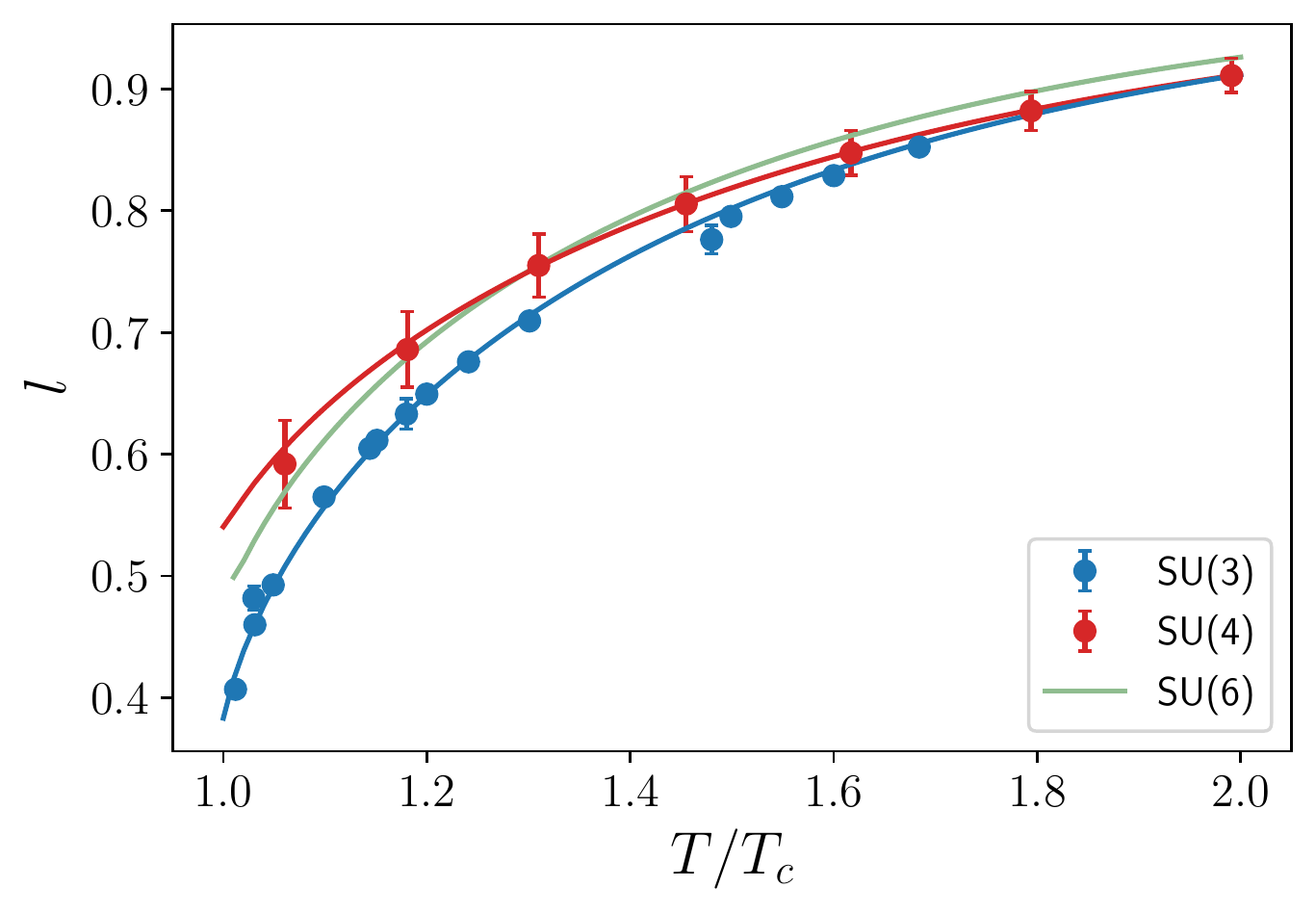}
    \caption{} \label{fig:fitsuN3}
\end{subfigure}
}
\caption{Best-fit to lattice data for the interaction measure (i) and Polyakov loop (ii) obtained from the model with a nonperturbative potential given in Eq. (\ref{eqn:suN_Vnpt}) for $N=3, 4$ and $6$ colors.}
\label{fig:fitsuN}
\end{figure}

\subsection{$G_2$ and $F_4$}
\label{sub:g2f4}

Now we generalize the discussion from the previous section to the exceptional groups $G_2$ and $F_4$. First, these groups have trivial centers and, also, the set of all roots for the lie algebras $\mathfrak{g}_2$ and $\mathfrak{f}_4$ are now divided into two sets, of long roots and of short roots, which do no mix under Weyl reflections. Both these facts combined decrease the amount of symmetry that can be imposed in the structure of the effective potential and, as a consequence, more terms are allowed in its construction.

Starting with $G_2$, the positive simple roots can be written as a linear combination of the weights in the lowest-dimensional representation\footnote{In the following, we refer to the lowest-dimensional representation as ``the fundamental''.} (\textbf{7}) as $\alpha_1 = \mu_1 - \mu_2$ and $\alpha_2 = - \mu_1$ (with $\langle \alpha_1, \alpha_1 \rangle >\langle \alpha_2, \alpha_2 \rangle$ and $\mu_1$, $\mu_2$ are weights in the fundamental), so that a general element of the Cartan subalgebra is given by, in the fundamental representation,
\begin{equation}
H = \alpha_1(H) \,{\rm diag}(0,-1,1, 0,1,-1,0) + \alpha_2(H) \,{\rm diag}(-1,-1,2, 1,1,-2,0) \equiv q_i H_i~,
\end{equation}

\noindent again with $q_i \equiv \alpha_i(H)$. Note that a trivial center requires the matrix exponentials $\exp(2 \pi \mathrm{i} H_i)$ to be the identity, so the diagonal entries of the matrices $H_i$ have to be integer numbers. When written in terms of the positive simple roots, the sets of positive long and short roots are\footnote{The set of all roots written as linear combinations of the simple roots are usually displayed as Hasse diagrams of the root poset of the Lie algebra.}, respectively, $\alpha_L = W \cdot \alpha_1 = \{\alpha_1, \alpha_1 + 3 \alpha_2, 2\alpha_1 + 3 \alpha_2 \}$ and $\alpha_S = W \cdot \alpha_2 = \{\alpha_2, \alpha_1 + \alpha_2, \alpha_1 + 2\alpha_2 \}$. Thus, the possible terms of the form (\ref{eqn:VsuN_terms}) can now have a sum running on either one of these sets of roots, i.e.,
\begin{align}
\begin{split}\label{eqn:g2f4_terms}
V^L_1 &= \frac{1}{2}\sum_{\alpha \in \alpha_L}  B_2(\alpha) ~, ~ V^L_2 = \frac{1}{8} \sum_{\substack{\alpha \neq \alpha^\prime  \\ \alpha, \alpha^\prime \in \alpha_L}} B_2(\alpha) B_2(\alpha^\prime) ~, ~ V^L_3 = \frac{1}{2} \sum_{\alpha \in \alpha_L} B_4(\alpha) \\
V^S_1 &= \frac{1}{2}\sum_{\alpha \in \alpha_S} B_2(\alpha) ~, ~ V_2^S = \frac{1}{8} \sum_{\substack{\alpha \neq \alpha^\prime  \\ \alpha, \alpha^\prime \in \alpha_S}} B_2(\alpha) B_2(\alpha^\prime) ~, ~ V^S_3 = \frac{1}{2} \sum_{\alpha \in \alpha_S} B_4(\alpha)
\end{split}
\end{align}

\noindent are the building blocks for the effective potential. The nonperturbative polynomial contribution to the effective potential can then be written as
\begin{align}
\begin{split}\label{eqn:V_g2f4}
V_{\rm npt}(q) = & ~T_{\rm c}^2 T^2 \left(c_0  + \sum_{i = 1}^3 c^L_i V^L_i (q)+ \sum_{i = 1}^3 c^S_i V^S_i (q) + c^{LS} V^L_1(q) V^S_1 (q)\right)
\\& + d_1 T_{\rm c}^4 + d_2 \frac{T_{\rm c}^6}{T^2}~.
\end{split}
\end{align}

The boundary of a Weyl chamber of $\mathfrak{g}_2$ is defined by the vanishing of the Killing form with the two positive simple roots, which are of different lengths, i.e., a point on the boundary obeys $\langle \alpha_i, H \rangle = 0$ for $i = 1$ or $2$. Therefore, the root system lacks the symmetry necessary for an assumption similar to the uniform eigenvalue ansatz, adopted in the previous section. Therefore, the effective potential $V(q_1, q_2)$ is necessarily two-dimensional; even if we impose that, initially, the global minimum of the potential lies equidistant from each hyperplane at the boundary of the Weyl chamber, the subsequent dynamics violates such condition.

\begin{figure}
\centering
\includegraphics[scale=0.65]{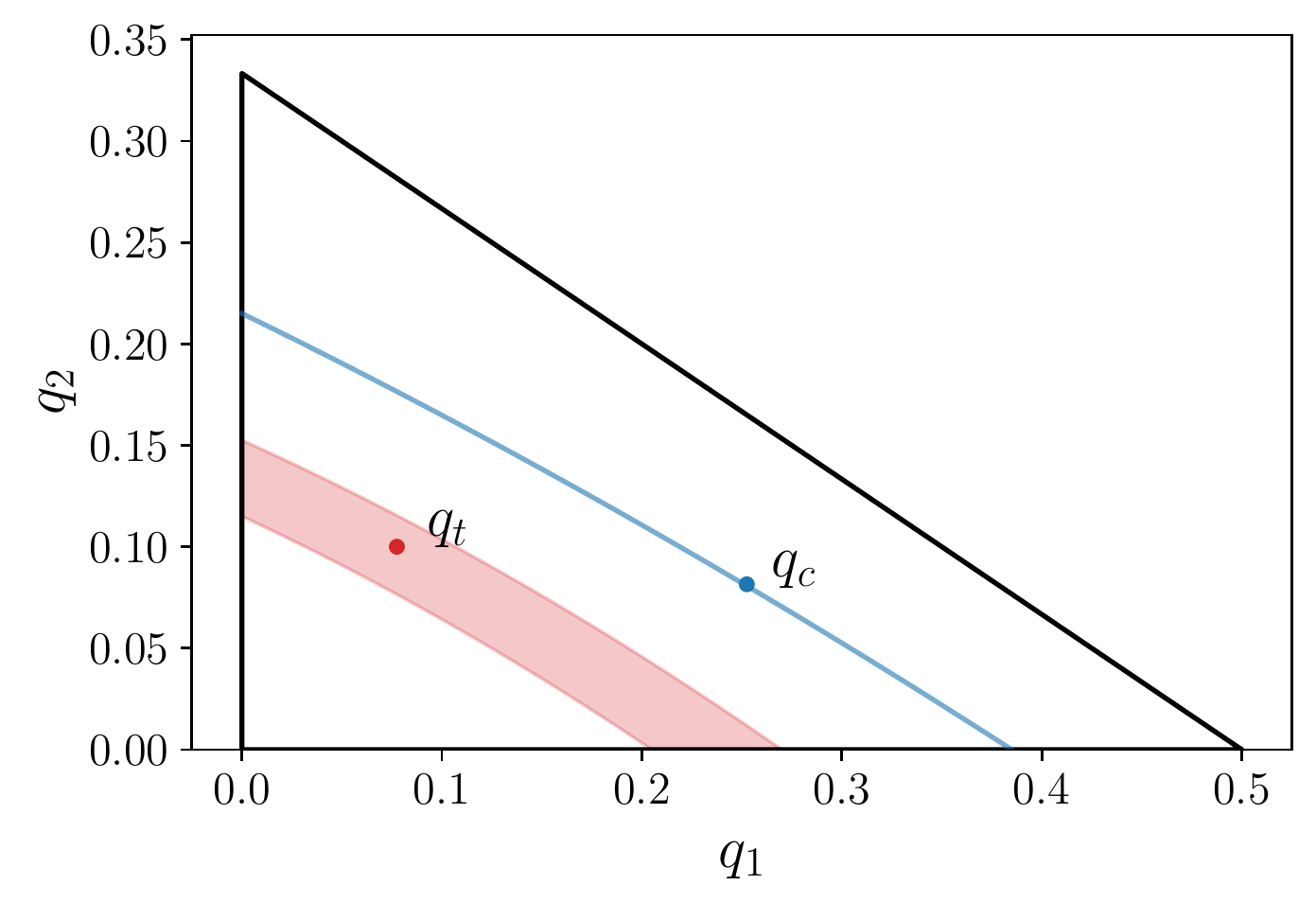}
\caption{A Weyl chamber for $G_2$ showing the coordinates for states with vanishing fundamental Polyakov loop (blue line) and for states with Polyakov loop in the range $0.38 \lesssim l(q_t) \lesssim 0.55$, corresponding to the expected region right above the critical temperature (red area). The two points show examples of randomly drawn states $q_c$ and $q_t$.}
\label{fig:g2chamber}
\end{figure}

The confined state, as in the case of SU($N$), is seen on the lattice \cite{PhysRevD.80.065028, PhysRevD.83.114502} to have a very small value of the traced Polyakov loop in the fundamental representation, $\langle l(T_{\rm c}^-) \rangle \ll 1$, which we take to vanish identically. Note that, as opposed to the case of SU($N$), this order parameter does not necessarily vanish below the critical temperature, as center symmetry is absent\footnote{In this context, confinement can be viewed as a consequence of the repulsion between the eigenvalues of the thermal Wilson line at low temperatures, instead of a direct consequence of center symmetry. \cite{Poppitz:2012nz, Anber:2014lba, Dunne:2016nmc}}. A priori, any element of the Cartan subalgebra $\mathfrak{h}$ with vanishing Polyakov loop can be taken as the confined state. This set defines a line in the Cartan subalgebra of $\mathfrak{g}_2$ on which we allow the confined state to be located, shown in blue in Fig. \ref{fig:g2chamber} along with the interior of a Weyl chamber.

Having constructed the potential, as the sum of Eq. (\ref{eqn:Vpt}) and (\ref{eqn:V_g2f4}), we then proceed as done in the case of SU($N$) and impose the following constraints. First, at $T = T_{\rm c}$ the global minimum of the potential jumps discontinuously, as the temperature is raised, from the confined state $q_c$ to a state with coordinates $q_t$, both with (approximately) vanishing pressure at that temperature. As mentioned previously, $q_c$ is randomly chosen subject to the condition $l(q_c) = 0$ and, based on the behavior of lattice data for SU($N$), we choose (also drawing randomly) the state $q_t$ so that $0.38 \lesssim l(q_t) \lesssim 0.55$ (this interval is taken from the expected region of Fig. \ref{fig:ls}). The region inside the Weyl chamber in Fig. \ref{fig:g2chamber} that obeys this bound is shown in red. Once both $q_c$ and $q_t$ are chosen, these conditions reduce the ten coefficients in Eq. (\ref{eqn:V_g2f4}) to four, which are then fitted by the lattice data (only available in the case of $G_2$) or expected lattice behavior. Specifically, for the observables not yet calculated on the lattice, we fit the model to the midsection of the expected regions in Figs. \ref{fig:deltas} and \ref{fig:ls}. We also selected the potentials that give values for the pressure that are as close as possible to zero in a temperature range $\delta T \sim 0.1 T_{\rm c}$ right below the critical temperature. This is imposed in an attempt to extrapolate the model to temperatures slightly below $T_{\rm c}$, so that the gravitational wave signal can be reliably calculated.

A similar construction can be made for the group $F_4$. Its positive simple roots can be written as \cite{F4roots, yokota2009exceptional}
\begin{equation}\label{eqn:f4_mus}
\left(\begin{array}{c}
\alpha_1 \\
\alpha_2 \\
\alpha_3 \\
\alpha_4
\end{array}\right) = \left(\begin{array}{cccc}
0 & 1 & -1 & 0 \\
0 & 0 & 1 & -1 \\ 
0 & 0 & 0 & 1 \\ 
1/2 & -1/2 & -1/2 & -1/2
\end{array}\right) \left(\begin{array}{c}
\mu_1 \\
\mu_2 \\
\mu_3 \\
\mu_4
\end{array}\right)~,
\end{equation}

\noindent which can be inverted to give the weights $\mu_i$ as a function of the roots $\alpha_i$. One can then write a general element $H \in \mathfrak{h}$, in the fundamental representation of the Lie algebra, as
\begin{align}
\begin{split}
H = &~\alpha_1(H) \,{\rm diag} (1, 1, 0, 0, ...) + \alpha_2(H) \,{\rm diag} (2, 1, 1, 0, ...) + \,\alpha_3(H) \,{\rm diag} (3, 1, 1, 1, ...)\\& + \alpha_4(H) \,{\rm diag} (2, 0, 0, 0, ...) \equiv q_i H_i~,
\end{split}
\end{align}

\noindent again with $q_i \equiv \alpha_i(H)$ and the basis $\{H_1, H_2, H_3, H_4\}$ having integer elements on the diagonal (the numbers after the ellipsis are determined by writing the additional weights $\mu$ as linear combinations of the positive simple roots\footnote{Note that, as in the case of $\mathfrak{g}_2$, the weights in the fundamental of $\mathfrak{f}_4$ are the short roots.}). One can then again divide the roots into sets of long roots $\alpha_L$ and short roots $\alpha_S$ and construct the possible terms in the nonperturbative potential as in Eq. (\ref{eqn:g2f4_terms}).

The boundaries of the Weyl chamber are now defined by the vanishing of the Killing form with respect to the four simple roots, two of which are long ($\alpha_1$ and $\alpha_2$) and the other two short ($\alpha_3$ and $\alpha_4$). The potential should be invariant under the Weyl group, which includes transformations that permute each pair (long or short) of positive simple roots. This allows for a simplifying assumption generalizing the uniform eigenvalue ansatz described in the case of SU($N$): we can take the minima of the potential to be always located at the plane that is equidistant from the Weyl chamber boundary hyperplane defined by the two long roots and also from the hyperplane defined by the short roots\footnote{This is also observed to hold only approximately in the exact solution, deviations are large close to the critical temperature.}. In other words, we can project the potential to the plane defined by $q_1 = q_2$ and $q_3 = q_4$, reducing the dimensionality of the effective potential from four to two. We emphasize that this is not a necessary assumption. However, both in the case of the uniform eigenvalue ansatz for SU($N$) as well as for its generalized version in the case of $F_4$, the model can accurately fit the (expected) behavior of thermodynamic quantities from lattice, so hopefully not much is lost by our assumption.

\begin{figure}
\makebox[1\textwidth][c]{
\begin{subfigure}{0.6\textwidth}
    \centering\includegraphics[scale=0.65]{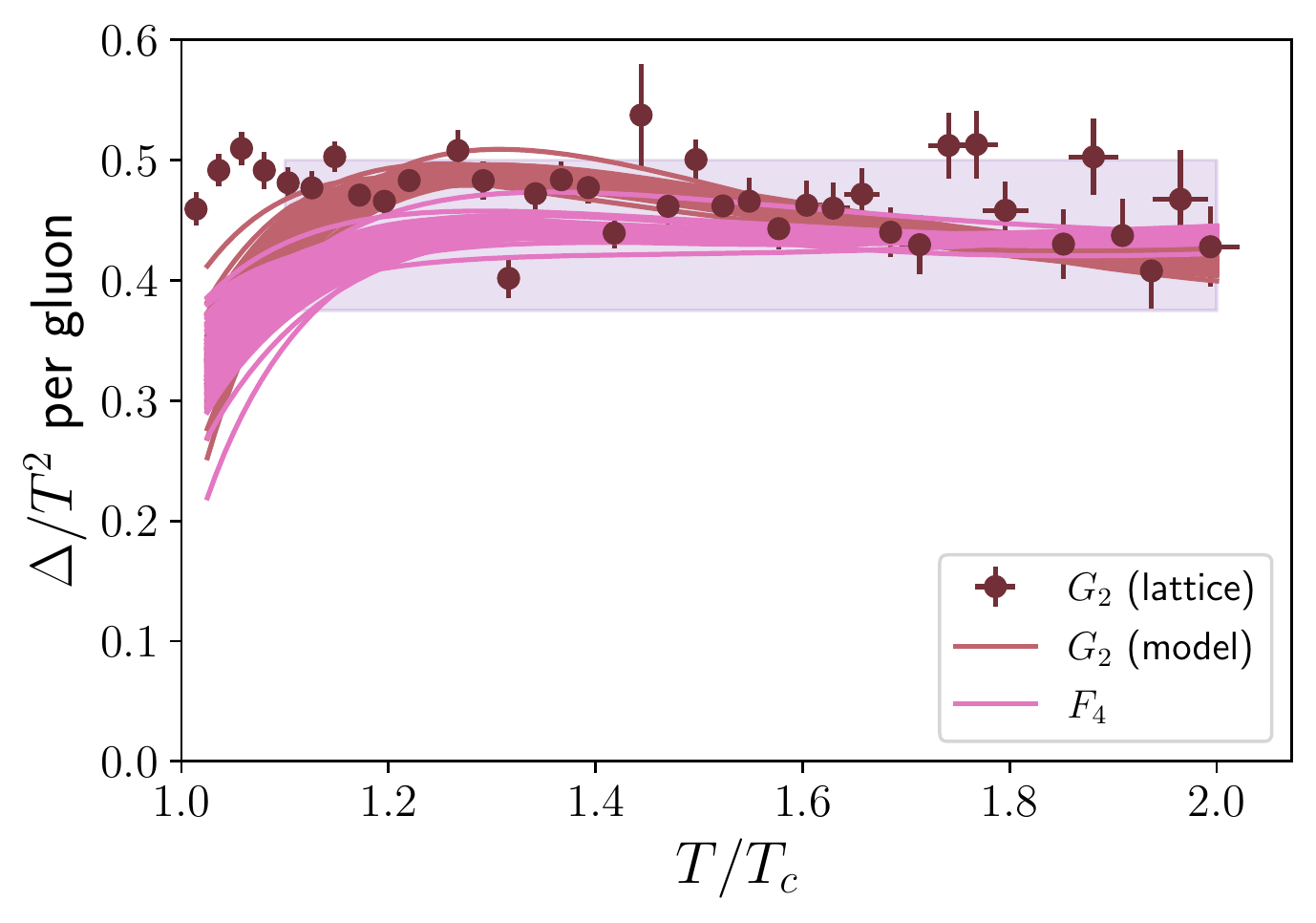}
    \caption[a]{} \label{fig:g2f4thermo1}
\end{subfigure}
\begin{subfigure}{0.6\textwidth}
    \centering\includegraphics[scale=0.65]{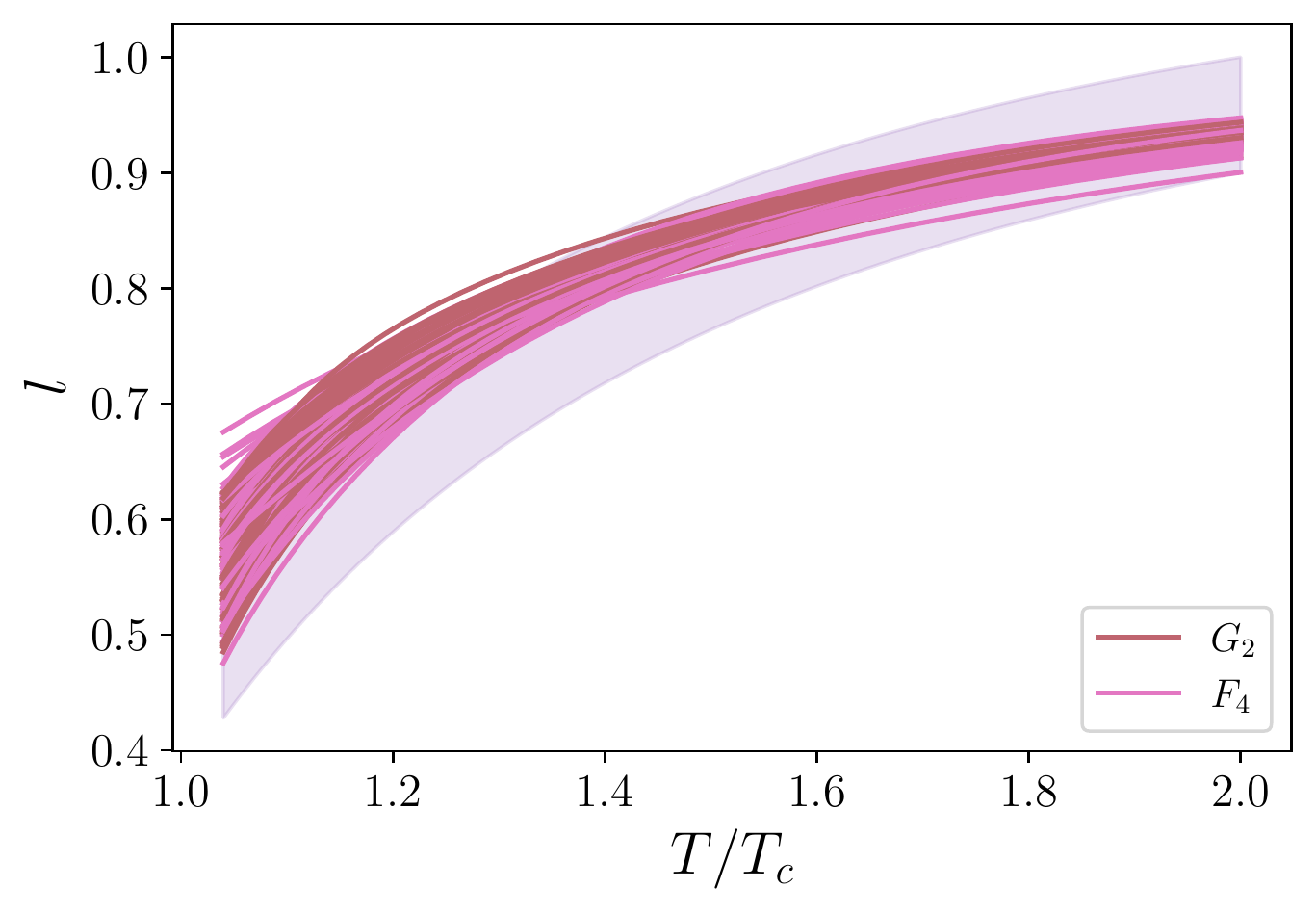}
    \caption{} \label{fig:g2f4thermo2}
\end{subfigure}
}
\caption{Resulting fits for the model given in Eq. (\ref{eqn:V_g2f4}) in the case of $G_2$ and $F_4$ gauge groups, showing the interaction measure and Polyakov loop curves obtained for different choices of $q_c$ and $q_t$ that best reproduce the lattice data available for $G_2$ or the expected regions shown in purple.}
\label{fig:g2f4thermo}
\end{figure}

The resulting curves for the fits to the interaction measure and the Polyakov loops for both $G_2$ and $F_4$ are presented in Fig. \ref{fig:g2f4thermo}. These plots show the resulting best-fit curves for many different choices of confined state $q_c$ and $q_t$. As the figure shows, we were able to construct a number of effective potentials that reproduce the (expected) lattice behavior.

\section{Stochastic gravitational wave signal}
\label{sec:gws}

Equipped with effective potentials for the semi-QGP phase in pure Yang-Mills theories, in this Section we compute stochastic gravitational wave spectra produced during the associated confinement transitions and study their possible observation in planned experiments.

First-order phase transitions in the early universe are well-known sources of a stochastic gravitational wave background \cite{Witten:1984rs, Kamionkowski:1993fg, Croon_2018}. This type of transition proceeds via nucleation and subsequent expansion of bubbles of the true ground state of the theory on a background in the metastable vacuum. There are different mechanisms that can generate gravitational radiation during a first-order PT (for a recent detailed description, see the reviews \cite{Caprini:2015zlo, Caprini:2019egz}); first, gravitational waves are produced during the collision of the expanding bubbles and, subsequently, the energy released to the thermal plasma by the transition generate sound wave and magnetohydrodynamic turbulence contributions. The scalar field contribution from bubble collisions is subdominant in the case of a nonrunaway PT, in which the bubble wall reaches a finite terminal velocity due to friction exerted by the thermal plasma. In that case, the fraction of the latent heat that becomes kinetic energy of the scalar field is vanishingly small, which renders the contribution of bubble collisions to the stochastic GW signal negligible.

In the case of a confining PT in a dark gauge sector without matter, i.e. dark Yang-Mills, the order parameter is the Polyakov loop, which is constructed out of the temporal component of the non-Abelian vector potential (see Eq. (\ref{eqn:polyloop})). Thus, the scalar field generating the bubbles of true vacuum should interact strongly with the thermal plasma surrounding them. Therefore, we assume that the confinement phase transition proceeds via nonrunaway bubbles and only account for the contribution of sound waves and turbulence to the stochastic background of gravitational waves. This is in agreement with the results of Ref. \cite{Bodeker:2017cim}, which shows that transition splitting, radiation emitted from gauge bosons acquiring a mass when traveling across the bubble wall from symmetric to broken phase, generates enough friction to impede the runaway of the bubble. In a confining PT, gauge bosons go from a deconfined to a bound state when crossing the bubble wall, and it is thus reasonable to expected that enough friction is generated by the plasma and the phase transition should follow the nonrunaway case.

\subsection{Parameters of the phase transition}
\label{sub:params}

Before calculating the spectrum of gravitational waves from the first-order PT, a number of parameters, particular to each physical model, have to be determined. 

First, the strength of the PT is encoded in the parameter $\alpha$, determined by the ratio of the change in the interaction measure $\Delta$ across the phase transition to the total thermal energy density of the universe in the symmetric phase, given in terms of the enthalpy $w \equiv e + p$, as
\begin{equation}\label{eqn:alpha}
\alpha = \frac{\Delta (T_{\rm n}^+) - \Delta (T_{\rm n}^-)}{3 w(T_{\rm n}^+)}~,
\end{equation}

\noindent calculated at the nucleation temperature $T_{\rm n}$, where the signs refer to the symmetric ($+$) and broken ($-$) phases. This temperature is the one at which there is on average one bubble of the confined phase nucleated per Hubble volume, which implies
\begin{equation}
\frac{S_3(T)}{T} \biggr \vert_{T = T_{\rm n}} = 2 \log \left( \frac{90 M_{\rm Pl}^2}{g_* \pi^2 T_{\rm n}^2}\right)~,
\end{equation}

\noindent with $M_{\rm Pl}$ being the reduced Planck mass and $g_*$ is the number of relativistic degrees of freedom at $T = T_{\rm n}$. For small amounts of supercooling $T_{\rm n} \lesssim T_{\rm c}$ (which turns out to be the case for the transitions considered here), the nucleation temperature is also very close to the bubble percolation temperature $T_*$ at which the PT can successfully complete and GWs are produced. Henceforth, we take $T_{\rm c} \approx T_{\rm n} \approx T_*$.

Another important parameter is the inverse duration of the PT, defined as
\begin{equation}\label{eqn:beta}
\frac{\beta}{H_*} = T_* \frac{d(S_3/T)}{dT} \biggr \vert_{T = T_*}~,
\end{equation}

\noindent with $H_*$ the Hubble parameter at $T_*$ and $S_3$ is the action for the O(3)-symmetric bounce solution for a thermal transition between the metastable and the true vacua. In the case of pure Yang-Mills described in Section \ref{sec:Vs}, this action can be obtained by going one step further in the operator expansion, adding the gauge kinetic term at leading order, as
\begin{equation}\label{eqn:S3}
S_3 = \int d\Omega \rho^2 d\rho \left( \frac{1}{2} {\rm tr} \mathbf{F^2_{\mu \nu}} + V(q) \right)~,
\end{equation}

\noindent where $\rho$ and $\Omega$ are, respectively, the three-dimensional radial coordinate and solid angle and $\mathbf{F_{\mu\nu}} \equiv F^a_{\mu\nu}T^a$. 

In addition to $\alpha$ and $\beta$, one should also determine the bubble wall velocity $v_{\rm w}$ as well as the efficiency factors $\kappa_{\rm v}(v_{\rm w}, \alpha^{\rm d})$ and $\kappa_{\rm tb}(v_{\rm w}, \alpha^{\rm d})$ for conversion of latent heat into bulk and turbulent motion, respectively. A proper determination of the bubble wall velocity necessitates a treatment of the dynamics of the bubble expansion, with an appropriate modeling of the friction terms (see, e.g. \cite{Baldes:2020kam} for a discussion of effects that contribute to this dynamics). This is, however, outside of the scope of this work and we assume a relativistic bubble wall velocity $v_{\rm w} \simeq 1$, expected to hold for values of $\alpha$ not much smaller than $\mathcal{O}(1)$. On the other hand, the efficiency factors are not additional parameters, as they depend exclusively on the bubble wall velocity and on
\begin{equation}\label{eq:alpha_h}
\alpha^{\rm d} \equiv \frac{\Delta (T_{\rm n}^+) - \Delta (T_{\rm n}^-)}{3 w^{\rm d}(T_{\rm n}^+)} = \frac{w(T_{\rm n}^+)}{w^{\rm d}(T_{\rm n}^+)} \alpha
\end{equation}

\noindent with $w^{\rm d}$ denoting the enthalpy in the dark sector only. In the limit $v_{\rm w} \simeq 1$, one has \cite{Espinosa_2010}
\begin{equation}
\kappa_{\rm v} \simeq \frac{\alpha^{\rm d}}{0.73 + 0.083 \sqrt{\alpha^{\rm d}} + \alpha^{\rm d}}~.
\end{equation}

\noindent Note that, if the dark sector dominates the energy density of the universe at the time of GW production, we get $\alpha \approx \alpha^{\rm d}$ and a single parameter suffices. Moreover, if the phase transition is fast enough, i.e., $\beta/H_* \gg 1$, which we show to be the case for the pure Yang-Mills confining transition in the following sections, bulk motion quickly becomes turbulent and one can take $\kappa_{\rm tb} \simeq \kappa_{\rm v}$  \cite{Ellis:2019oqb}. 

\subsection{Energy budget and glueball-dominated phase}
\label{sub:budget}

We now address some cosmological considerations that maximize the GW signal.

For fixed values of the parameters discussed in the previous section, the energy density in the form of gravitational waves $\rho_{\rm gw, *}$ emitted during the PT in a given dark sector is directly proportional to the radiation energy density in that sector $\rho_{\rm rad, *}^{\rm d}$ prior to confinement. If the universe is radiation-dominated at the time of the transition, the GW density parameter right after production obeys
\begin{equation}\label{eq:omegastar}
\Omega_{\rm gw, *} \sim \frac{\rho_{\rm gw, *}}{\rho_{\rm rad, *}} = \frac{\rho_{\rm gw, *}}{\rho_{\rm rad, *}^{\rm d} + \rho_{\rm rad, *}^{\rm other}}~,
\end{equation}

\noindent with $\rho_{\rm rad, *}^{\rm other}$ being the radiation energy density in other sectors (which include, of course, the visible sector). With the ratio $\rho_{\rm gw, *}/\rho_{\rm rad, *}^{\rm d}$ fixed, a maximal signal is obtained if the dark sector going though the confinement phase transition dominates the energy density of the universe at the time of the transition, i.e., for $\rho_{\rm rad, *}^{\rm d} \gg \rho_{\rm rad, *}^{\rm other}$, which we assume from here on\footnote{For cases in which the dark sector is cold with respect to the visible sector or when many sectors contribute to the total energy density, see \cite{Breitbach_2019, Fairbairn_2019} and \cite{Archer_Smith_2020}, respectively.}.

As mentioned in Section \ref{sub:params}, if most of the energy density is in the dark sector at the time of the PT, one has $\alpha \approx \alpha^{\rm d}$. Now, lattice results show that the pressure of the gluon gas in the semi-QGP is negligible below $T_{\rm c}$ and that the pressure is continuous across the confinement PT, so the parameter $\alpha$ is reduced to
\begin{equation}
\alpha \approx \alpha^{\rm d} \approx \frac{e (T_{\rm n}^+) - e (T_{\rm n}^-)}{3 e(T_{\rm n}^+)} \approx \frac{e(T_{\rm c}^+)}{3 e(T_{\rm c}^+)} = \frac{1}{3}~,
\end{equation}

\noindent as the energy density in the confined state is small compared to its value above the critical temperature and $T_{\rm n} \approx T_{\rm c}$. In general, if the energy density in other sectors cannot be neglected, we get $\alpha < \alpha^{\rm d} \approx 1/3$ as $w_{\rm d} (T_n^+) \leq w (T_n^+)$ in Eq. (\ref{eq:alpha_h}). This decrease in the value of $\alpha$ further suppresses the GW signal when compared to the case in which the dark sector is dominant.

The density parameter of gravitational waves redshifted to today, $\Omega_{\rm gw}$, depends on the detailed evolution of the Hubble parameter since the time of GW production. In particular, if a confining dark sector dominates the energy density at the time of the PT, one expects to have a period of matter domination\footnote{That is not exactly true, as glueball $3 \rightarrow 2$ self-interactions, while still active, make them redshift slightly faster than matter; the correction factor is, however, a slowly varying logarithm in the scale factor, i.e., $\propto \log(a)$ \cite{Carlson:1992fn, Halverson:2016nfq, Halverson:2018olu}.} after confinement occurs, with most of the energy density of the universe in the form of dark glueballs, $\rho_{\rm gb, *} \approx \rho_{\rm rad, *}^{\rm d}$. If that happens before big bang nucleosynthesis (BBN), the glueballs ultimately have to decay (mostly) to radiation in the visible sector before the onset of BBN, as a persistent early matter domination phase would spoil its predictions. For simplicity, we assume that glueballs decay directly to visible sector radiation at some later time\footnote{We assume an instantaneous decay of the glueballs, as that is sufficient to estimate the order of magnitude of the entropy exchanged between the sectors. For a more careful treatment, see e.g. \cite{Jo:2020ggs}.}, when the scale factor is $a_{\rm \tau}$. During their lifetime, the energy density in glueballs increases as $\propto a$ relative to the GW energy density, so that at the time of decay
\begin{equation}
\frac{\rho_{\rm gw, \tau}}{\rho_{\rm gb, \tau}} = \frac{a_*}{a_{\rm \tau}} \frac{\rho_{\rm gw, *}}{\rho_{\rm gb, *}} \approx \frac{a_*}{a_{\rm \tau}} \frac{\rho_{\rm gw, *}}{\rho_{\rm rad, *}^{\rm d}}~.
\end{equation}

\noindent with $a_*$ being the scale factor at the time of bubble percolation. Then, the energy density in glueballs is transferred to visible sector radiation, so that right after their decay $\rho_{\rm rad, \tau}^{\rm v} = \rho_{\rm gb, \tau}$, and the density parameter in GWs becomes
\begin{equation}\label{eq:omegatau}
\Omega_{\rm gw, \tau} \sim \frac{\rho_{\rm gw, \tau}}{\rho_{\rm rad, \tau}^{\rm v}} ~\Rightarrow~ \Omega_{\rm gw, \tau} \approx \frac{a_*}{a_{\rm \tau}} \Omega_{\rm gw, *}~,
\end{equation}

\noindent where we used Eq. (\ref{eq:omegastar}) in the limit $\rho_{\rm rad, *}^{\rm d} \gg \rho_{\rm rad, *}^{\rm other}$. Thus, a longer period of early matter domination means a stronger suppression of the GW signal\footnote{For more on the effect of matter domination on the gravitational wave signal, see \cite{Barenboim_2016}.}, compared to the case in which the dark glueballs decay to visible sector radiation almost immediately after the PT.

We estimate the maximum amplitude of the GW spectrum by assuming that the glueballs decay quickly to the visible sector, in such a way that the factor $a_*/a_{\rm \tau}$ in Eq. (\ref{eq:omegatau}) is approximately one. This amounts to a situation in which the lifetime of glueballs with respect to SM decays is much shorter than the age of the universe at BBN, $T_{\rm BBN} \sim \mathcal{O}(\rm min)$. Given the requirement of gauge symmetry, the lowest dimension operator connecting the dark and visible sectors is of dimension six \cite{Forestell:2017wov}, of the form
\begin{equation}\label{eqn:decayop}
    \mathcal{L} \supset \frac{1}{M^2} H^\dagger H G^a_{\mu\nu} G^{a,\mu\nu}~,
\end{equation}

\noindent with $H$ the SM Higgs doublet and $M$ being the mass scale of the degrees of freedom connecting visible and dark sectors\footnote{Such an operator can be generated by integrating out either scalar and fermionic mediators with masses $\sim M$ that couple to the SM Higgs and are charged under the dark gauge group, see \cite{Forestell:2017wov}.}. For a confinement scale\footnote{This value is taken here for the dark confinement scale since it is the one that maximizes the projected reach of the future GW searches, i.e., BBO and DECIGO.} of $\Lambda \sim 100$ GeV, the lifetime of glueballs is smaller than $1$s for $M \lesssim 10^8$ GeV (see, e.g., Fig. 2 of \cite{Forestell:2017wov}). Therefore, we assume the presence of the higher-dimensional operator in Eq. (\ref{eqn:decayop}) with $\Lambda \sim 100$ GeV $ << M \lesssim 10^8$ GeV, so that dark glueballs decay quickly enough and our description of the dark sector as pure Yang-Mills is justified.

Given the assumptions above, of instantaneous confinement transition and glueball decay with negligible lifetime, we can relate the the temperature of the visible sector plasma right after glueballs decay (which also coincides with bubble percolation), $T_*^{\rm v}$, to the confinement scale in the dark sector $\Lambda_{\rm d} \lesssim T_{\rm c}$. The energy density originally in dark radiation is ultimately transformed into energy in the visible sector plasma, so that
\begin{equation}\label{eq:Tstarv}
g^{\rm v}_* (T_*^{\rm v})^4 \approx g_*^{\rm d} T_*^4  ~\Rightarrow~ T_*^{\rm v} \approx \left(\frac{g_*^{\rm d}}{g_*^{\rm v}}\right)^{1/4} T_* \sim \left(\frac{g_*^{\rm d}}{g_*^{\rm v}}\right)^{1/4} \Lambda_{\rm d}~,
\end{equation}

\noindent with $g^{\rm d}_*$ and $g^{\rm v}_*$ the number of relativistic degrees of freedom in the dark and visible sectors, respectively, at percolation. For finite glueball lifetimes, the temperature $T_{\rm gb}^{\rm v}$ in the visible sector right after the decay is given by Eq. (\ref{eq:Tstarv}) multiplied by the factor $(a_*/a_\tau)^{3/4}$.

\subsection{Energy density in gravitational waves}
\label{sub:spectra}

As discussed at the beginning of this section, gravitational waves in a nonrunaway PT are produced both by sound waves and turbulence, the former giving a larger contribution. The total energy density produced is the sum of the terms \cite{Caprini:2015zlo, Hindmarsh:2017gnf, Caprini:2019egz}
\begin{align}\label{eqn:GWsw}
\frac{d \Omega_{\rm sw}}{ d \log(f)} &= 0.687 F_{\rm gw} K^{3/2} \left(\frac{H_* R_*}{\sqrt{c_{\rm s}}} \right)^2 \tilde{\Omega}_{\rm gw} S_{\rm sw} (f)\\\label{eqn:GWtb}
h^2 \frac{d \Omega_{\rm tb}}{ d \log(f)} &= 3.20 F_{\rm gw} K^{3/2} (H_* R_*)  S_{\rm tb}(f)~,
\end{align}

\noindent with $R_* = (8 \pi)^{1/3} v_{\rm w}/\beta$ being the mean bubble separation at percolation, $c_{\rm s} \sim 1/\sqrt{20}$ \cite{Heinz:2005ja} the speed of sound in the plasma at $T_c$, $\tilde{\Omega}_{\rm gw} \sim 10^{-2}$ a numerical factor obtained from simulations and $K$ the fraction of kinetic energy in the plasma, given by 
\begin{equation}
K = \frac{\kappa_{\rm v} \alpha}{1 + \alpha} ~.
\end{equation}
\noindent Note that the expression for the sound wave contribution in Eq. (\ref{eqn:GWsw}) already takes into account the suppression factor for short-lasting PTs, recently discussed, e.g., in \cite{Guo:2020grp}. In addition, the spectral shape functions have the form
\begin{align}
S_{\rm sw}(f) &= \left(\frac{f}{f_{\rm sw,0}} \right)^3 \left(\frac{7}{4 + 3f^2/f_{\rm sw,0}^2} \right)^{7/2}\\
S_{\rm tb}(f) &= \left(\frac{f}{f_{\rm tb,0}} \right)^3\frac{(1+f/f_{\rm tb,0})^{-11/3}}{ 1 + 8 \pi f/h_*}~,
\end{align} 

\noindent with $h_* = a_*H_*/a_0$ the inverse Hubble time at percolation redshifted to today and the the peaks are at the frequencies (redshifted to today)
\begin{align}\label{eqn:peakfreq}
f_{\rm sw,0} &\approx 26 \left(\frac{1}{H_*R_*} \right) \left(\frac{z_{\rm p}}{10} \right) \left(\frac{T_*^{\rm v}}{ 100 ~\rm GeV} \right) \left(\frac{g_*^{\rm v}}{100}\right)^{1/6}~\mu{\rm Hz}\\
f_{\rm tb,0} &\approx 79 \left(\frac{1}{H_* R_*} \right) \left(\frac{T_*^{\rm v}}{ 100 ~\rm GeV} \right) \left(\frac{g_*^{\rm v}}{100}\right)^{1/6}~\mu{\rm Hz}~,
\end{align}

\noindent with $z_p \simeq 10$ obtained in numerical simulations.

The remaining factor in Eqs. (\ref{eqn:GWsw}) and (\ref{eqn:GWtb}), $F_{\rm gw}$, accounts for the redshift of the amplitude of the GW density parameters from the time of emission to today, being therefore sensitive to assumptions about the intermediate cosmic evolution of the universe. As discussed in Section \ref{sub:budget}, we neglect the lifetime of the glueballs produced in the confining transition, so that the universe follows the standard cosmic evolution after glueball decay and
\begin{equation}
F_{\rm gw} = \Omega_{\gamma,0} \left(\frac{g^{\rm v}_{\rm s,0}}{g^{\rm v}_{\rm s, *}} \right)^{4/3} \frac{g^{\rm v}_*}{g^{\rm v}_0}
\end{equation}

\noindent with $\Omega_{\gamma,0}$ the energy density in photons today and $g^{\rm v}_{s, *}$, $g^{\rm v}_{s, 0}$ the effective numbers of entropic degrees of freedom in the visible sector at after glueball decay and today, respectively, and $g^{\rm v}_0$ is the number of relativistic degrees of freedom today.

Note that the expressions in Eqs. (\ref{eqn:GWsw}) and (\ref{eqn:GWtb}) are valid for sources that are not long-lasting, i.e., for $\beta/H_* \gg \mathcal{O} (1)$. To calculate the sound wave contribution, we use the web-based tool \textsf{PTPlot} introduced in \cite{Caprini:2019egz}.

\subsection{Results}
\label{sub:gwresults}

\begin{figure}
\makebox[1\textwidth][c]{
\begin{subfigure}{0.6\textwidth}
    \centering\includegraphics[scale=0.65]{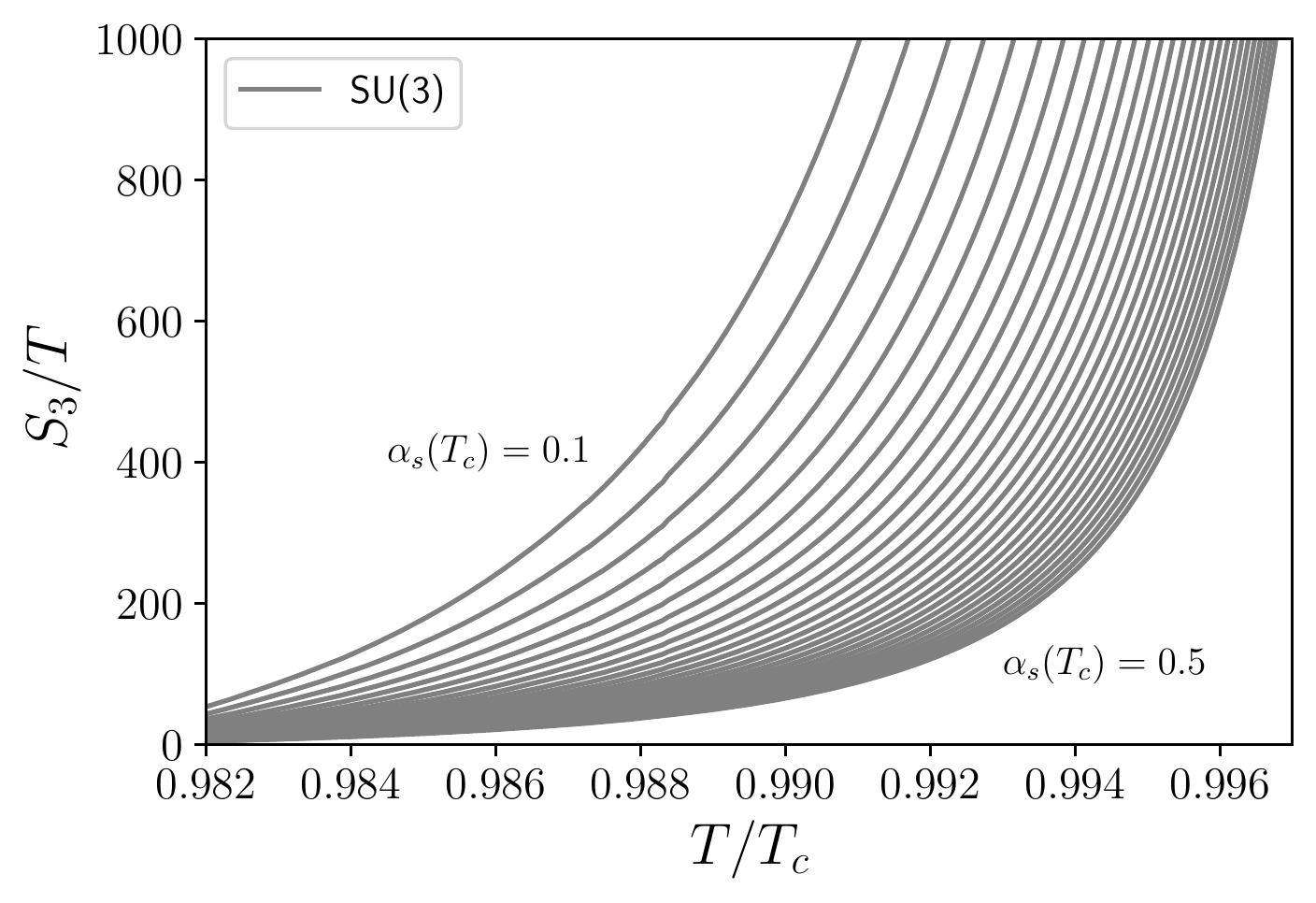}
    \caption[a]{} \label{fig:su3Ss}
\end{subfigure}
\begin{subfigure}{0.6\textwidth}
    \centering\includegraphics[scale=0.65]{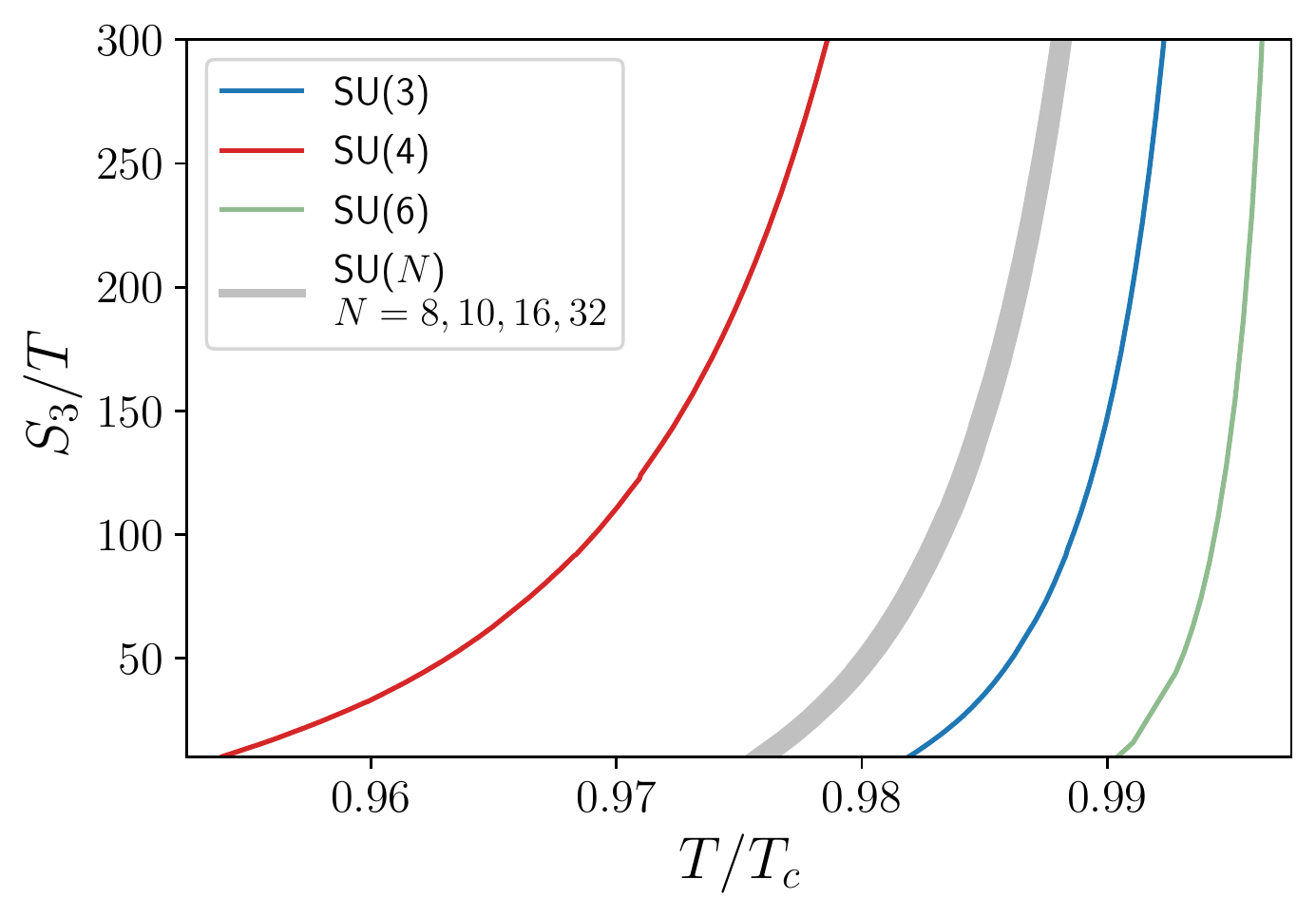}
    \caption{} \label{fig:actions}
\end{subfigure}
}
\caption{The action for the bounce solution as a function of temperature (i) in the case of SU(3) for different choices of $\alpha_s(T_c)$ and (ii) for different numbers of colors . For $N \geq 8$, the model saturates to the curve shown in gray, expected from Eqs. (\ref{eqn:VsuN_terms_r}) and (\ref{eqn:latheat}) as explained in the text.}
\label{fig:actionsall}
\end{figure}

Before using the \textsf{CosmoTransitions} package, the action in Eq. (\ref{eqn:S3}) has to be written in terms of canonically normalized fields $\vec \varphi$ as
\begin{equation}\label{eqn:suNaction_can}
S_3 = \int d\Omega \rho^2 d\rho \left[ \frac{1}{2}\left(\frac{d \vec \varphi}{d \rho} \right)^2 + V[q(\vec \varphi)] \right]~.
\end{equation}
\noindent For example, for the gauge groups discussed in Section \ref{sec:Vs}, these fields can be written written as 
\begin{equation}
\begin{dcases}
\varphi = \left(\frac{\pi T}{g} \sqrt{\frac{2(N^2-1)}{3 N}} \right) r  & \text{for SU(}N)\\
\varphi_1 = \frac{2 \sqrt{2} \pi T}{g} q_1 ~, ~\varphi_2 = \frac{4 \sqrt{6} \pi T}{g} \left(\frac{q_1}{2} + q_2 \right) & \text{for} ~G_2\\
\varphi_1 = \frac{4 \pi T}{g} \sqrt{\frac{51}{7}} q_1 ~, ~\varphi_2 = \frac{4 \pi T}{g} \left(9 \sqrt{\frac{3}{7}}q_1 + 2 \sqrt{21} q_3 \right) & \text{for} ~F_4
\end{dcases}~,
\end{equation}

\noindent where the variable $r$ introduced in the discussion above Eq. (\ref{eqn:VsuN_terms_r}). The canonically normalized fields for SU($N$) without the uniform eigenvalue ansatz can also be easily obtained, but are not shown explicitly here. The action $S_3$ as a function of temperature can then be calculated for each value of the dark coupling constant $\alpha_{\rm s}(T_{\rm c}) = g^2/4 \pi$ at the critical temperature, which we assume to be in the interval $\alpha_{\rm s}(T_{\rm c}) \in (0.2, 0.4)$ for all gauge groups\footnote{This choice is motivated by the value $\alpha_{\rm s}(T_{\rm c}) \sim 0.3$ for SU($3$) and SU($4$) on the lattice \cite{Gockeler:2005rv, Lucini:2008vi}.}. Examples are shown in Fig. \ref{fig:actionsall} for SU($N$). Fig. \ref{fig:su3Ss} shows how the action $S_3$ changes with the choice of $\alpha_s(T_c)$ and \ref{fig:actions} shows $S_3$ for $N = 3$ (in blue), $4$ (red), $6$ (green) and  $8, 10, 16, 32$ (all falling on the gray band) with $\alpha_{\rm s}(T_{\rm c}) = 0.3$. For $N < 8$, we find the action $S_3$ with the full potential, without the uniform eigenvalue assumption, while for $N \geq 8$ that we use the simplified one-dimensional potential, using Eq. (\ref{eqn:VsuN_terms_r}).

From these results for the action $S_3$ as a function of temperature, one can determine the values of the inverse duration $\beta$ from Eq. (\ref{eqn:beta}) for the confining transition in each case, the result being shown in Table \ref{table:1}. Note that for all numbers of colors $ \geq 8$, the curves for the action $S_3(T)$ as a function of temperature are approximately identical, so that the value of $\beta$ should not vary significantly for large values of $N$. This is to be expected; the $N$-dependent terms in Eqs. (\ref{eqn:VsuN_terms_r}) and (\ref{eqn:latheat}) that cannot be absorbed by the free coefficients in Eq. (\ref{eqn:suN_Vnpt}) are all $\mathcal{O} (1/N^2)$. As a consequence, each term in Eq. (\ref{eqn:VsuN_terms_r}) is (up to normalization) approximately independent of $N$ for large $N$.

\begin{table}[h!]
\centering
\begin{tabular}{ |c | c | }
\hline
 	&  $\beta/H* $  \\
 SU(3) & $(5.1 \pm 0.6) \times 10^4$\\ 
 SU(4) & $(2.9 \pm 0.6) \times 10^4$ \\  
 SU(6) & $(7.7 \pm 2.2) \times 10^4$   \\
 SU(8)  & $(4.0 \pm 0.8) \times 10^4$\\
\hline 
\end{tabular}
\caption{Values of the parameter $\beta/H_*$ for SU($N$). Errors are estimated by calculating $\beta$ with different coefficients $c_i$ and $d_i$ which also give good fits to lattice data.}
\label{table:1}
\end{table}

\begin{figure}
\makebox[1\textwidth][c]{
\begin{subfigure}{0.6\textwidth}
    \centering\includegraphics[scale=0.65]{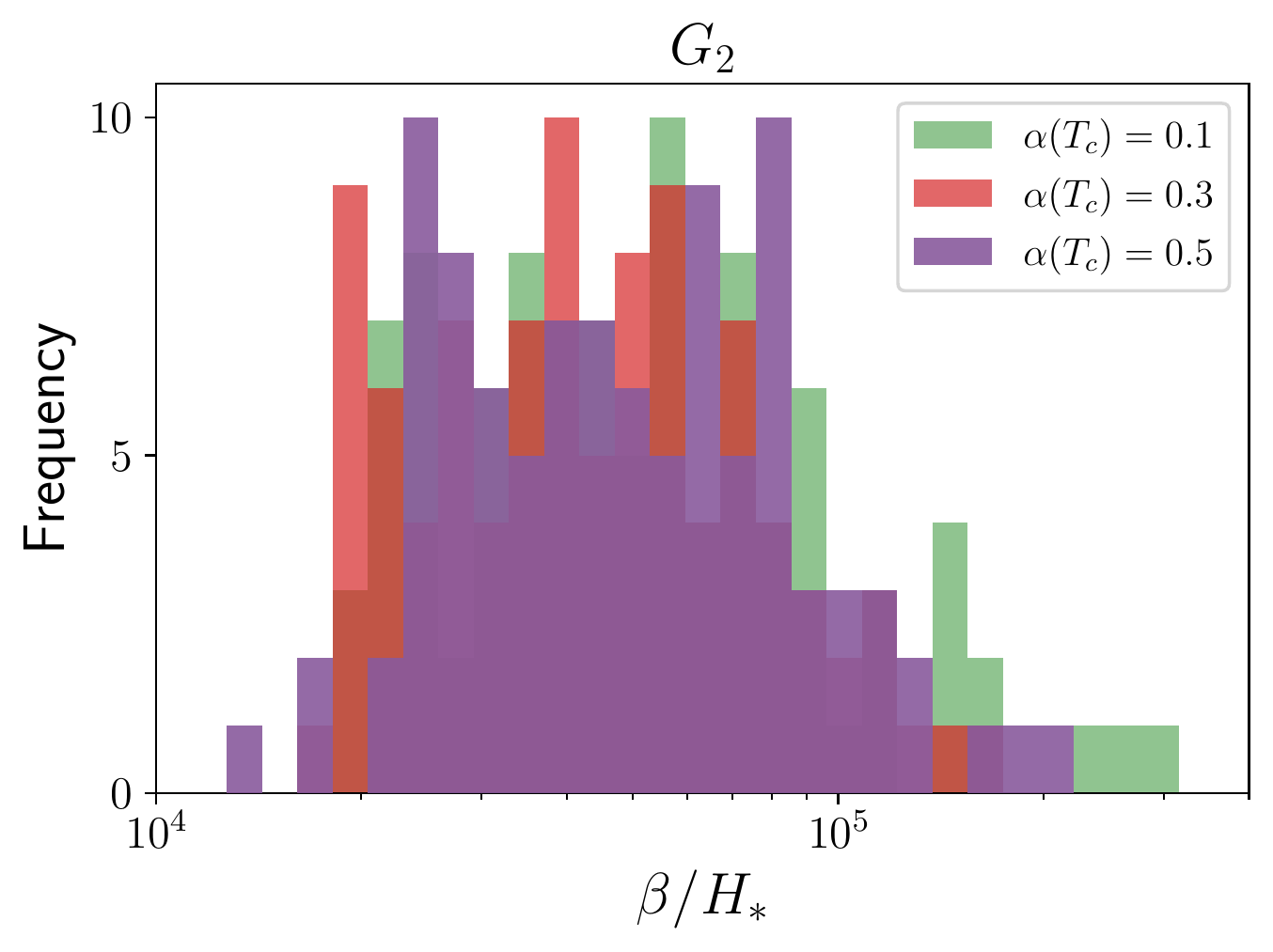}
    \caption[a]{} \label{fig:betasg2}
\end{subfigure}
\begin{subfigure}{0.6\textwidth}
    \centering\includegraphics[scale=0.65]{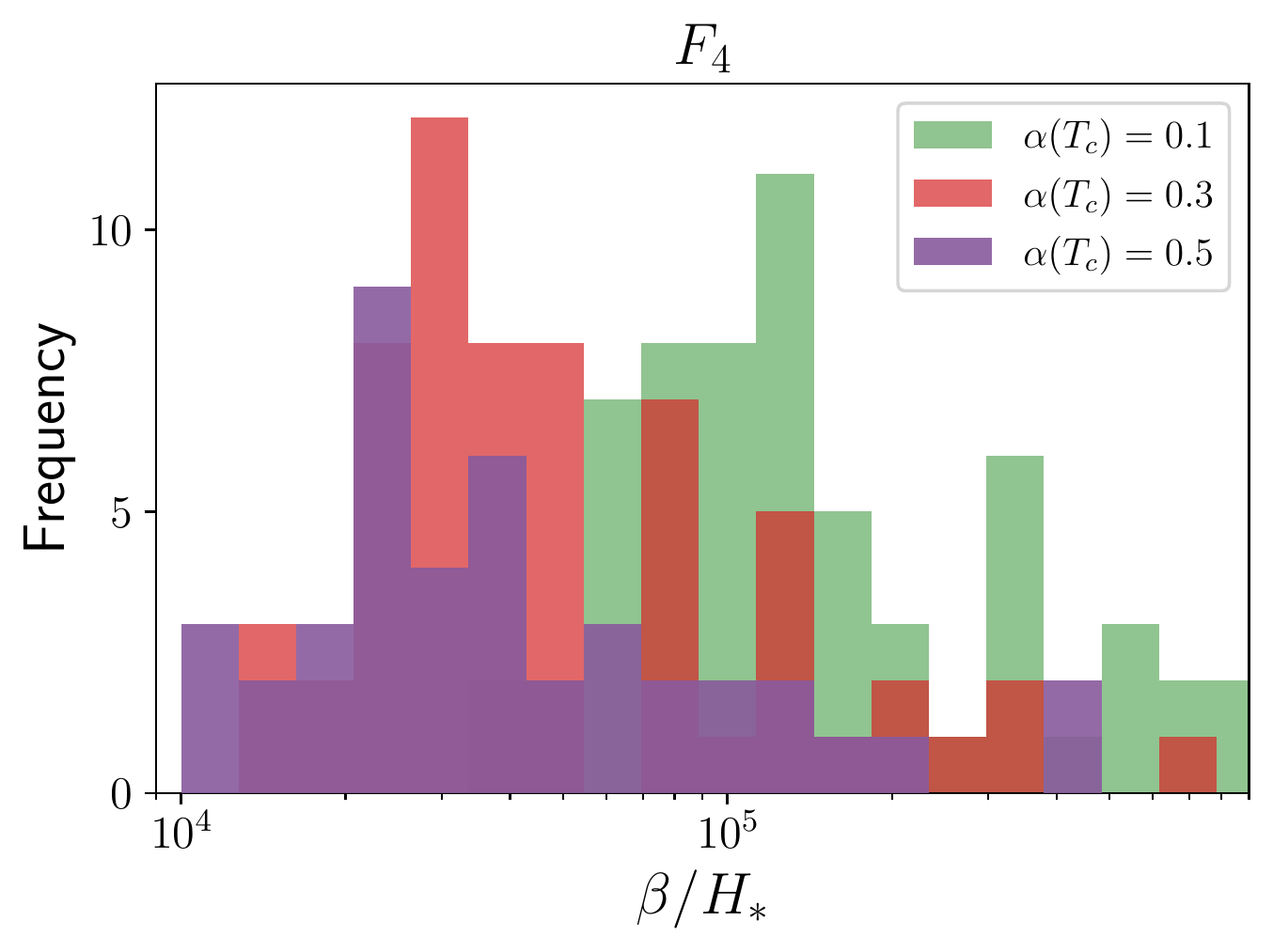}
    \caption{} \label{fig:betasf4}
\end{subfigure}
}
\caption{The parameter $\beta$ giving the inverse duration of the confining PT in the case of $G_2$ (i) and $F_4$ (ii). For most effective potentials used in the calculation for both these gauge groups, the result was in the range $\beta/H_* \sim 10^4 - 10^5$, with a similar distribution for the three choices of coupling $\alpha(T_{\rm c}) = 0.1, 0.3$ and $0.5$ for $G_2$ and a slight dependence of the choice of coupling for $F_4$.}
\label{fig:g2f4betas}
\end{figure}

For $G_2$ and $F_4$, the value of the parameter $\beta$ depends on the choice of states $q_c$ and $q_t$,  defined in Section \ref{sec:Vs}. For the models shown in Fig. \ref{fig:g2f4thermo}, the distribution of values of $\beta$ is shown in Figs. \ref{fig:betasg2} and \ref{fig:betasf4} for the values of the dark coupling constant $\alpha(T_{\rm c}) = 0.1, 0.3$ and $0.5$. These distributions seem to be independent of the value of the coupling for $G_2$ and to have a slight dependence on the coupling for $F_4$, with peaks at $\beta/H_* \sim 10^4$ in all cases except $F_4$ with $\alpha(T_{\rm c}) = 0.1$, which peaks at $\beta/H_* \sim 10^5$.  Similar values of $\beta/H_*$ were also observed in other effective models, e.g., describing the chiral phase transition in confining dark sectors with matter \cite{Helmboldt_2019, Aoki_2020}. This gives a GW signal orders of magnitude smaller than estimated with more optimistic choices for the duration of the PT, e.g., in \cite{Schwaller:2015tja}. (See \cite{Bigazzi:2020avc, GarciaGarcia:2016xgv} for related works that also discuss gravitational waves from confining PTs.)

\begin{figure}[h!]
\makebox[1\textwidth][c]{
\begin{subfigure}{0.65\textwidth}
    \centering\includegraphics[scale=0.65]{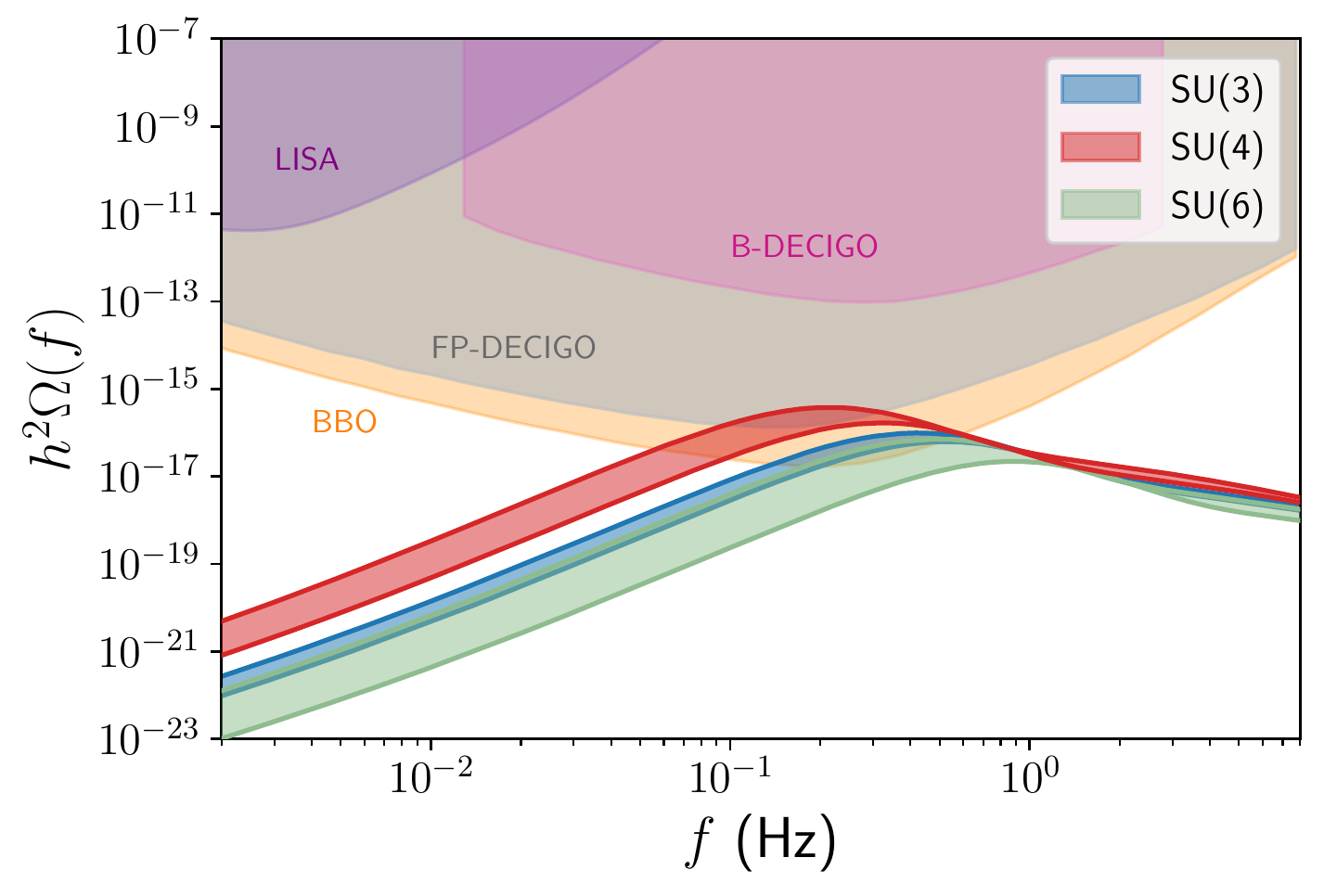}
    \caption{} \label{fig:gwssuN}
\end{subfigure}
\begin{subfigure}{0.65\textwidth}
    \centering\includegraphics[scale=0.65]{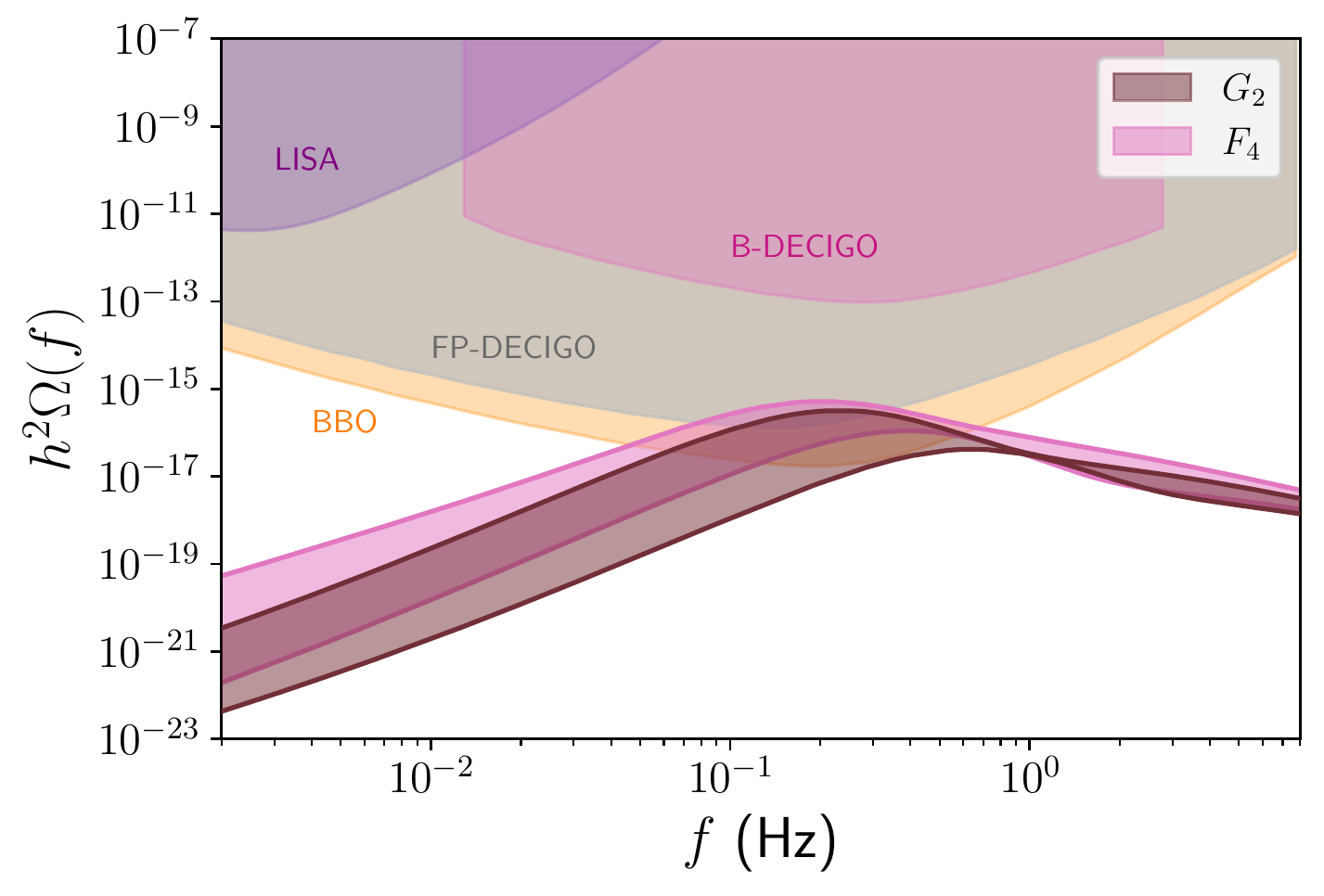}
    \caption{} \label{fig:gwsg2f4}
\end{subfigure}
}
\caption{The resulting GW spectra for (i) SU($N$) with $N = 3,4,6$ and (ii) $G_2$, $F_4$ for the choice $T_*^{\rm v} = 100$ GeV, $g_*^{\rm v} = 100$. The critical temperature in each case can be obtained from Eq. (\ref{eq:Tstarv}) as $T_{\rm c} \approx 158, 135, 109, 137$ and $99$ GeV repectively for SU(3), SU(4), SU(6), $G_2$ and $F_4$. This is of the same order as the dark confinement scale $T_{\rm c} \sim \Lambda$, although the exact relationship depends on the gauge group and requires lattice calculations (see footnote \ref{foot}). The uncertainty comes mostly from the different choices of good fits for SU($N$) and of states $q_c$ and $q_t$ in the case of $G_2$ and $F_4$. The value of the dark coupling constant at the critical temperature is taken to be in the range $0.2 \leq \alpha_{\rm s}(T_{\rm c}) \leq 0.4$, which also contributes to the uncertainty.}
\label{fig:gws}
\end{figure}

Once the parameter $\beta$ is determined, the gravitational wave signal can be calculated from Eqs. (\ref{eqn:GWsw}) and (\ref{eqn:GWtb}). The resulting spectra for $T_*^{\rm v} = 100$ GeV and $g_*^{\rm v} = 100$ (the approximate number of relativistic degrees of freedom in the standard cosmic evolution at $T_*^{\rm v}$) are shown in Fig. \ref{fig:gws}, along with projected experimental sensitivities for next-generation (LISA), taken from \textsf{PTPlot}, and next-to-next generation (BBO and DECIGO) gravitational wave searches, adapted from Refs. \cite{Seto_2001, Kawamura:2006up, Crowder_2005, Harry:2006fi, Brdar:2018num}. The shaded strips represent the uncertainty estimated by varying the best-fit coefficients of the model in Eqs. (\ref{eqn:suN_Vnpt}) and (\ref{eqn:V_g2f4}) as well as different choices of $q_c$ and $q_t$ in the case of $G_2$ and $F_4$. Additional uncertainty comes from varying the dark coupling constant in the interval $\alpha_{\rm s}(T_{\rm c}) \in (0.2, 0.4)$. The signal is within range of next-to-next generation searches if $T_*^{\rm v} \sim 100$ GeV, although many orders of magnitude out-of-reach of LISA. For different values of $T_*^{\rm v}$, the maximum amplitude of the signal changes only slightly (due to a small change in $T_{\rm n}$); from Eq. (\ref{eqn:GWsw}), the energy density of GWs can be seen to depend on the combination $H_* R_* \sim H_*/\beta$, which is independent of $T_*^{\rm v}$ in our model. On the other hand, the peak frequency varies linearly with $T_*^{\rm v}$ (Eq. (\ref{eqn:peakfreq})), so that any significant deviation from $T_*^{\rm v} \sim 100$ GeV pushes the signal out of the observable range.

\section{Summary and conclusions}
\label{sec:concl}

In this work we studied stochastic gravitational wave backgrounds produced by confining phase transitions in dark Yang-Mills sectors. This requires constraining effective potentials by symmetry and lattice considerations, constructing them for concrete simple Lie groups, and then computing the gravitational wave signal. We describe each in turn.

In Section \ref{sec:constraints} we set the stage for constructing an effective matrix model for the semi quark-gluon plasma in pure Yang-Mills theories. To do so, we discuss the necessary symmetries (center symmetry and Weyl group invariance) as well as lattice observables that constrain the effective potential on thermal Wilson line eigenvalues that take values in the Cartan subalgebra and serve as order parameters for the phase transition.

In Section \ref{sec:Vs}, these constraints were implemented, yielding (for each group) an effective potential that models the behavior of a strongly coupled gas of gluons close to the confinement phase transition. Such behavior is determined by currently available lattice data for SU($N$) gauge groups with small numbers of colors ($N=3,4,6$) as well as for $G_2$. The universality observed in this data was used to extend the matrix model to describe the exceptional gauge groups $F_4$, as well as SU($N$) with larger numbers of colors. Assuming universal thermodynamic behavior, we showed that a simple effective model can appropriately describe observable quantities such as the interaction measure and the renormalized Polyakov loop for all gauge groups considered.

Equipped with the effective potentials for the confinement transitions, we computed the stochastic gravitational wave background in Section \ref{sec:gws}. This requires the determination of the action of the bounce solution in a thermal transition between a confined state and a (partially) deconfined one, which in turn allowed for an estimation of the gravitational wave signal.
For all gauge groups considered, the GW signal is only accessible to futuristic experimental searches such as BBO and DECIGO, being many orders of magnitude below the projected reach of LISA. This happens because the PT is not long-lasting, having an inverse duration parameter $\beta/H_* \sim 10^4$ or larger, suppressing the GW energy density emitted by sound waves in the plasma. In addition, this signal is only visible when the glueballs resulting in a dark sector decay to visible sector radiation at a temperature $T_*^{\rm v} \sim 100$ GeV. For temperatures not of this order of magnitude, the spectrum's peak frequency takes the signal out of the range of observation of both BBO and DECIGO. Interestingly, $T_*^{\rm v}\sim 100$ GeV occurs when the dark confinement scale is near the weak scale.

Though this signal is relatively weak, its interest derives from the fact that  the only surefire model-independent way to detect dark sectors is gravitationally. As much as we might wish for stronger portals, they simply may not exist, a stubborn fact that is unfortunately consistent with all current evidence for dark sectors. However, in spite of these sobering facts, the importance of dark sectors simply demands a deeper understanding of gravitational probes, even when potential signals are decades away.

\vspace{1cm}
\noindent \textbf{Acknowledgments.} We thank Vinicius Aurichio, Yang Bai, Huaike Guo, Brandon Melcher, Manuel Reichert, Alastair Wickens, Susan van der Woude and especially Mustafa Amin, Kaloian Lozanov, and Scott Watson for useful conversations as well as Marco Panero for providing lattice data. G.S. is grateful to Frederico Campos Freitas for assistance with the Discovery Cluster. J.H. is supported by NSF CAREER grant PHY-1848089. The work of C.L. is supported in part by the Alfred P. Sloan Foundation Grant No. G-2019-12504. B.N. and G.S. are supported by NSF grant PHY-1913328.

\section*{Appendix}

In this Appendix, we show the values for the coefficients in the nonperturbative potentials of Eqs. (\ref{eqn:suN_Vnpt}) and (\ref{eqn:V_g2f4}) that give a best fit to the lattice data in \cite{Datta:2010sq, Panero:2009tv}. Table \ref{table:1} gives the parameters for the SU($N$) fits and Table \ref{table:2} for $G_2$ and $F_4$.

\begin{table}[h!]
\centering
\begin{tabular}{ |c | c | c | c | c | c |c |}
\hline
 	&  $c_0$ & $c_1$ & $c_2$ & $c_3$ & $d_1$ & $d_2$  \\
 SU(3) & 3.52 & -18.3 & 5.28 & 29.9 & 5.82 & -2.52  \\
 SU(4) & 8.90 & -19.0 & -0.78 & 34.1 & 45.2 & -1.67  \\  
 SU(6) & 19.0 & -18.7 & 1.45 & 34.1 & 14.6 & -5.41  \\
 SU(8)  & 36.6 & -18.9 & 0.61 & 38.2 & 20.0 & -6.68  \\
\hline 
\end{tabular}
\caption{Coefficients in Eq. (\ref{eqn:suN_Vnpt}) for SU($N$) best-fit curves.}
\label{table:1}
\end{table}

\begin{table}[h!]
\centering
\begin{tabular}{ | c | c | c | c | c | c | c | c | c | c | c |}
\hline
 	&  $c_0$ & $c^L_1$ & $c^L_2$ & $c^L_3$ & $c^S_1$ & $c^S_2$ & $c^S_3$ & $c^{LS}$ & $d_1$ & $d_2$\\
 $G_2$	& 1.79  & 18.6	& 9.20  & 3.71	&  -45.6 & 174	& 74.1  & -58.1	& 8.94  & -3.56 \\
$F_4$	&  16.4 &  -14.2 & 23.4  & 19.9	& -7.53  & 29.2	& -6.3  & -21.1	& 19.8  & -7.11 \\
\hline
\end{tabular}
\caption{Coefficients in Eq. (\ref{eqn:V_g2f4}) for $G_2$ and $F_4$ best-fit curves.}
\label{table:2}
\end{table}

\bibliography{refs}
\bibliographystyle{ieeetr}

\end{document}